\documentclass[prb,twocolumn,superscriptaddress,floatfix,showpacs]{revtex4}

\usepackage[dvips]{graphicx}
\usepackage{graphicx}
\usepackage{amsmath}
\usepackage{booktabs}
\usepackage{dcolumn}
\usepackage{textcomp}
\usepackage{rotating}

\newcommand{\lmo}{LaMnO$_3$}

\newcommand{\undoped}{LaSrMnO$_4$}

\newcommand{\laca}{La$_{1-x}$Ca$_{x}$MnO$_3$}

\newcommand{\ramo}{{\it R}$_{1-x}${\it A}$_{x}$MnO$_3$}

\newcommand{\ndsr}{Nd$_{1-x}$Sr$_{x}$MnO$_{3}$}

\newcommand{\ndsrhalb}{Nd$_{0.5}$Sr$_{0.5}$MnO$_{3}$}

\newcommand{\ndsrvier}{Nd$_{0.51}$Sr$_{0.49}$MnO$_{3}$}

\newcommand{\lsmov}{La$_{1-x}$Sr$_{x}$MnO$_{4}$}

\newcommand{\prnv}{Pr$_{0.6}$Ca$_{0.4}$MnO$_{3}$}

\newcommand{\lsmh}{La$_{0.5}$Sr$_{1.5}$MnO$_4$}

\newcommand{\tc}{T$_C$}
\newcommand{\tco}{T$_{CO}$}

\newcommand{\dx}{d$_{3x^2-r^2}$}
\newcommand{\dy}{d$_{3y^2-r^2}$}

\newcommand{\mnd}{Mn$^{3+}$}
\newcommand{\mnv}{Mn$^{4+}$}

\newcommand{\eg}{e$_g$}

\newcommand{\ef}{E$_f$}

\newcolumntype{d}{D{.}{.}{-1}}

\newlength{\figwidth}

\divide\figwidth by 2

\begin{document}

\advance\vsize by 2 cm

\title{Spin-wave excitations in the ferromagnetic-metallic and in the charge, orbital and spin ordered states in {\ndsr} with x$\approx$0.5}

\author{H. Ulbrich}
\email{ulbrich@ph2.uni-koeln.de}%
\affiliation{II. Physikalisches Institut, Universit\"at zu K\"oln, Z\"ulpicher Str. 77, D-50937 K\"oln, Germany}

\author{F. Kr\"uger}
\email{frank.kruger@st-andrews.ac.uk}%
\affiliation{School of Physics \& Astronomy, University of St. Andrews, North Haugh, St. Andrews, Fife KY16 9SS, United Kingdom}

\author{A. A. Nugroho}
\affiliation{Faculty of Mathematics and Natural Sciences, Institut Teknologi Bandung, Jl. Ganesha 10 Bandung, 40132 Indonesia}

\author{D. Lamago}
\affiliation{Forschungszentrum Karlsruhe, Institut für Festkörperphysik, P.O.B. 3640, D-76021 Karlsruhe, Germany} \affiliation{Laboratoire
L\'eon Brillouin, C.E.A./C.N.R.S., F-91191 Gif-sur-Yvette Cedex, France}

\author{Y. Sidis}
\affiliation{Laboratoire L\'eon Brillouin, C.E.A./C.N.R.S., F-91191 Gif-sur-Yvette Cedex, France}

\author{M. Braden}
\email{braden@ph2.uni-koeln.de}%
\affiliation{II. Physikalisches Institut, Universit\"at zu K\"oln, Z\"ulpicher Str. 77, D-50937 K\"oln, Germany}

\date{\today, \textbf{preprint}}

\pacs{PACS numbers: 75.10.Hk, 75.25.-j, 75.25.Dk, 75.30.Ds , 75.47.Lx, 75.40.Gb}

\begin{abstract}

We report inelastic neutron scattering experiments on single crystals of {\ndsr} with x=0.5 and x=0.49. The spin-wave dispersion in the charge,
orbital, and spin ordered state in {\ndsrhalb} exhibits a strongly anisotropic stiffness. The sign of the anisotropy is characteristic for the
site-centered model for charge and orbital ordering in half-doped manganites. Within this model, linear spin-wave theory yields a perfect
description of the experimental dispersion. In the ferromagnetic metallic state of {\ndsr} with x=0.49 and x=0.50 magnetic excitations exhibit
nearly the same magnon dispersion. High intense signals near the zone-boundary over a wide energy level overlap with a sharp spin-wave
dispersion which can be described with a Heisenberg model including nearest-neighbor interactions.

\end{abstract}

\maketitle

\section{Introduction}

In spite of its eminent relevance for colossal magnetoresistance (CMR) the exact nature of the charge, orbital and spin ordered (COSO) state in
the manganites \ramo ({\it R}=rare earth, {\it A}=bivalent alkaline earth) with an antiferromagnetic insulating ground state is still under
debate. The huge change in resistivity results from the competition between a ferromagnetic (FM) metallic and an antiferromagnetic (AFM)
insulating state. A comprehensive understanding of the magnetic ordering is essential to get a wide understanding of CMR.

The parent compound {\lmo} with a transition-metal valence of {\mnd} exhibits a ground state which is characterized by an antiferromagnetic
A-type ordering.\cite{jonk+50,sant+50} The ratio of {\mnd}:{\mnv} can be controlled by substitution of three-valent {\it R} with two-valent {\it
A} leading to complex phase diagrams with a correlated charge, orbital, and spin ordered ground state (see Ref. \onlinecite{Tokura+99} for a
brief summary). In spite of the numerous publications in perovskite (113) manganites around half-doping, its COSO state has not been fully
established. \cite{Radaelli+97,kawa+98,Brink+99,Yunoki+2000,Kajimoto+02} Some 113-manganites around x=0.5 exhibit an A-type AFM metallic or FM
metallic ground state while a decrease of the one-electron bandwidth leads to an AFM insulating state with CE-type ordering.\cite{kawa+98}

\begin{figure}
\centering
\includegraphics[width=6cm]{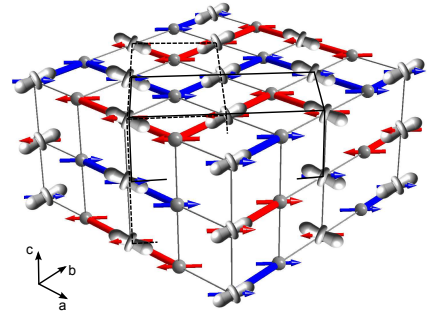}
\caption{(color online) Schematic representation of a half-doped perovskite manganite with charge, orbital and antiferromagnetic CE-type
ordering (Goodenough model). The pattern shows zigzag chains propagation in [1,1,0] direction (cubic notation). The {\eg} orbital ordering on
{\mnd} ions is shown by {\dx} and {\dy} orbitals. {\mnv} ions are drawn by balls while arrows stand for the magnetic alignment of the spins. The
orbital unit cell is indicated by dashed black lines and the magnetic unit cell of {\mnd} by black dotted lines.} \label{ce-pattern}
\end{figure}

The COSO state in half-doped {\laca} has been first studied by Wollan and Koehler and by Goodenough proposing a simple checkerboard ordering of
{\mnd} and {\mnv} ions (Goodenough model).\cite{Wollan+55,Gooden+55} Additionally, the {\eg} orbitals on {\mnd} sites order in stripe-like
patterns and spins order in an antiferromagnetic alignment called CE-type (see Fig. \ref{ce-pattern}). The magnetic moments follow the orbital
arrangement forming ferromagnetic zigzag chains while adjacent chains couple antiferromagnetically. First structural studies confirm the
Goodenough model qualitatively, but the charge modulation appears to be significantly below the ideal value. \cite{Radaelli+97, Jirak+00,
Herrero+04} Single-crystal neutron diffraction experiments on {\prnv} are not in agreement with the site-centered charge-ordering picture but
propose charge order on the Mn-O-Mn bonds, forming Zener polarons (Zener-polaron model) yielding ferromagnetically coupled
dimers.\cite{doud+02,Efremov+04} The nature of the COSO state is thus still matter of debate and a clear proof for one of these models is still
missing for \ramo .

The magnetic exchange coupling follows the orbital occupations. Consequently a careful study of these interaction parameters by analyzing the
spin-wave dispersion can give a direct view on the orbital arrangement. Indeed, for the single-layered (214) manganite {\lsmh}, inelastic
neutron scattering experiments established the spin-wave dispersion yielding strong support for the COSO state described qualitatively by the
Goodenough model.\cite{Senff+06} In the magnetic ground state, acoustical branches exhibit a steep spin-wave dispersion parallel to the zigzag
chains and a flatten dispersion perpendicular to them. These observations indicate a strong coupling inside the zigzag chains and a weak
coupling between adjacent chains as it is expected for the Goodenough model. On the contrary, the Zener-polaron model implies a strong
ferromagnetic coupling between the dimers and hence an anisotropy opposite to the observations.

Perovskite manganites exhibit intrinsical twinning rendering microscopic investigations quite complex. The high-temperature structural phase
transition associated with a lowering of the symmetry from cubic down to orthorhombic results in six different domain
orientations.\cite{Jung+05} The occurrence of charge and orbital order can only enhance the number of different domains. The magnetic neutron
scattering at a magnetic superstructure reflection in a twinned crystal is thus reduced by roughly a factor of six compared to a monodomain
crystal. The same holds for the inelastic signal arising  from a low-energy magnon. The interpretation of inelastic neutron scattering
experiments is further aggravated by unavoidable superposition of the signals from different domain orientations at higher energy.

Due to the absence of twinning, the spin-wave dispersion and detailed studies in the COSO state has first been studied in single-layered
manganites {\lsmov}.\cite{Reutler+03, Senff+05, Senff+06, Senff+08, Ulbrich+11} The {\eg}-orbitals in the undoped compound {\undoped} are
dominantly aligned perpendicular to the planes, but upon doping in {\lsmov} these {\eg}-orbitals flop into the MnO layers\cite{Senff+05}, so
that the ground state of the 214-manganites around half-doping qualitatively resembles that in 113-manganites at low
temperatures.\cite{Tokura+00} There is thus strong support, that the qualitative CE-model describes the ground-state in single-layered
half-doped {\lsmh}. However it remains an open issue whether the conventional AFM CE-type model may account for all COSO phases in manganites
with x$\approx$0.5, or whether the order essentially changes with the dimensionality (three-dimensional versus two-dimensional materials) and
with the structural distortions associated with the buckling of the MnO layers. Furthermore, in spite of its charge and orbital ordered
ground-state, single-layered manganites exhibit no CMR under moderate magnetic field most likely due to the enhanced stability of the COSO state
in layered materials.\cite{Tokunaga+99} Consequently, it is a matter of particular interest to extend the study of the spin-wave dispersion in
the COSO state to perovskite manganites. There is a tremendous number of publications on magnetic excitations in the ferromagnetic metallic
phase and in the A-type AFM phase in perovskite
manganites.\cite{kawa+98,Yoshizawa+98,Fernandez+98,Kawano-F+03,Endoh+05,Hennion+05,Kirby+06,Ye+06,Ye+07,Petit+09} But best to our knowledge,
investigations of the spin-wave dispersion in 113-manganites in the COSO state with AFM CE-type ordering are still missing. A microscopic
analysis of both, the ferromagnetic metallic and the antiferromagnetic insulating state is important to understand the switching between theses
phases by external parameters like pressure, magnetic field or temperature.

\begin{figure}
\includegraphics[width=0.5\textwidth]{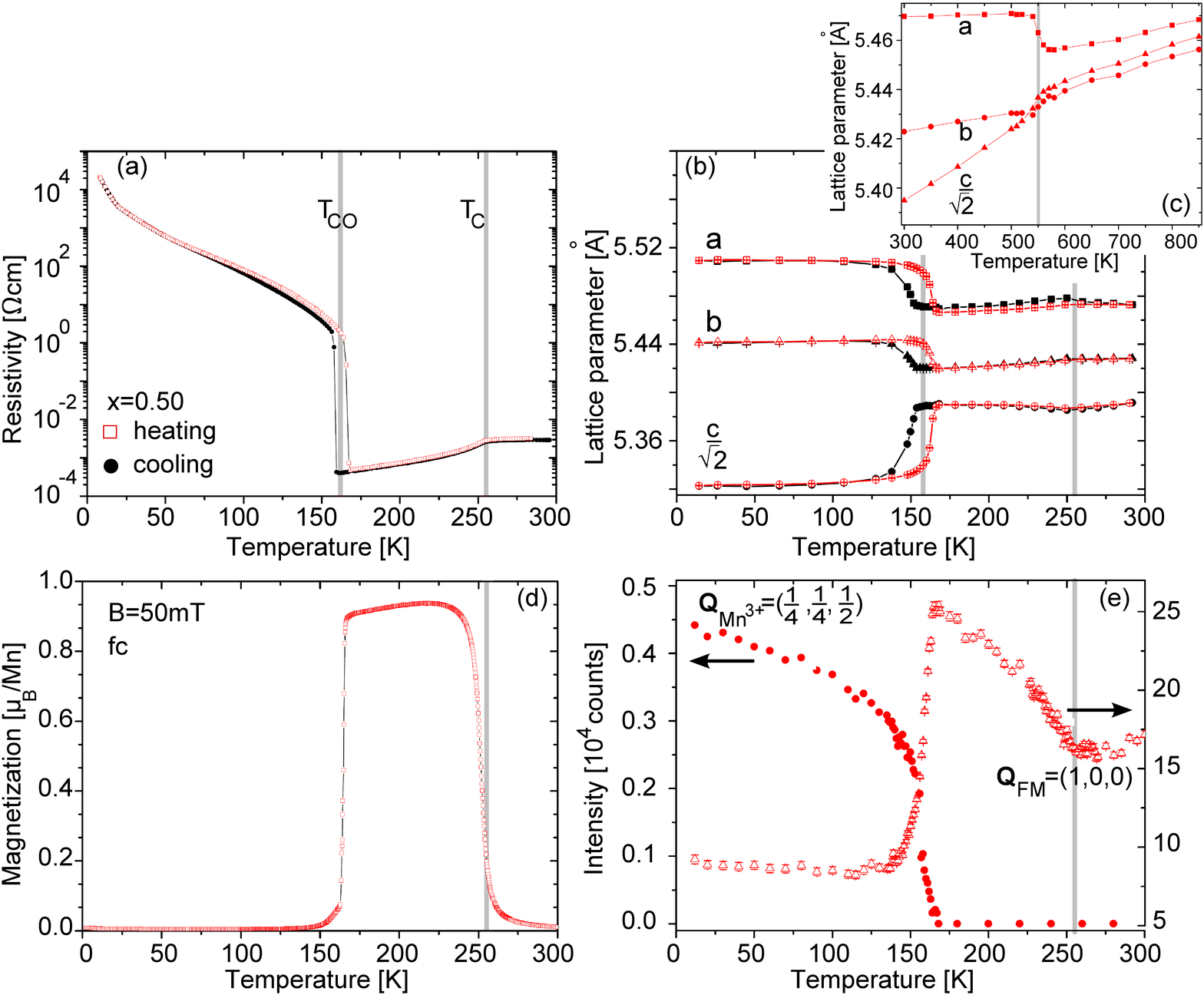}
\caption{(color online) Temperature-dependence of transport measurements and structural investigations of {\ndsrhalb}. Experimental data of
electric resistivity upon cooling show a kink at the transition into the (half-)metallic state at {\tc}=255 K (a), and the occurrence of the
COSO state at {\tco}$\approx$160 K causes a sudden jump in resistivity by three orders of magnitude pointing to insulating behavior.
Magnetization data indicating a ferromagnetically ordered state below {\tc} and an antiferromagnetically ordered state below {\tco} (d). X-ray
powder diffraction from T=15 K up to T=850 K shows remarkable changes of the orthorhombic structure at various temperatures in accordance with
electric and magnetic properties (b,c). Neutron diffraction at the superstructure reflection {\bf{Q}}$_{FM}$=(1,0,0) indicates ferromagnetic
ordering between {\tco}$<$T$<${\tc}. Furthermore, superstructure reflections referring to antiferromagnetic ordering of {\mnd} can be detected
below {\tco} (e). For more details see text.} \label{makr}
\end{figure}

In the paper we present experimental and theoretical analysis of the spin-wave dispersion in {\ndsr} with x=0.50 and x=0.49. Antiferromagnetic
CE-type ordering in {\ndsr} has been shown to occur only in a very small doping range near x=0.5.\cite{Kajimoto+99,Kajimoto+02,Tokura+06} A
slight change of stoichiometry from x=0.50 to x=0.49 can lead to a ground-state in which ferromagnetic ordering and antiferromagnetic CE-type
order coexist, while the ferromagnetic metallic (FMM) state above the COSO state is stable in a broader concentration range.\cite{Kajimoto+99}
The paper is divided as follows: After the experimental section we start by discussing the COSO state of the half-doped compound {\ndsrhalb}. We
present experimental data of the spin-wave dispersion up to energies of about 12 meV and compare these results with linear spin-wave theory. In
the second part we present inelastic scattering measurements in the FMM state of {\ndsrhalb} and of {\ndsrvier} along the main-symmetry
directions and give a description of the FM magnon dispersion by a simple Heisenberg model.

\section{Experimental}
Two large single crystals of {\ndsr} with x=0.49 and x=0.50 have been grown by the floating zone method. The starting materials,
Nd$_{2}$O$_{3}$, SrCO$_{3}$, and MnO$_{2}$ were preheated and mixed in appropriate stoichiometric proportions. After sintering for several times
at 1000$\,^{\circ}\mathrm{C} $ and $ 1200\,^{\circ}\mathrm{C} $ a cylindrical rod with 90 mm length and 10 mm diameter was pressed and heated at
1500$\,^{\circ}\mathrm{C} $. The apparatus used for crystal growth was the floating-zone furnace (Crystal Systems Incorporated), equipped with
four halogen lamps. The feed and seed rods were rotated in opposite directions at about 15 rpm. The molten zone was vertically moved at a rate
of 2-3 mm/h in a pressure of pure oxygen atmosphere (2 bar). The high quality of the two samples has been checked by macroscopic and microscopic
studies. Magnetization was measured by a SQUID magnetometer, electric resistivity by a standard four-contact method, and X-ray powder
diffraction experiments were carried out on a D5000 Siemens diffractometer.

{\ndsr} with x=0.49 and x=0.50 crystallizes in the orthorhombic unit-cell with space group Ibmm and room-temperature lattice parameters of
$a{\approx}\sqrt2{\times}a_{p}$, $b{\approx}\sqrt2{\times}a_{p}$, $c{\approx}2{\times}a_{p}$, where $a_{p}$ is the lattice parameter of the
cubic perovskite. In the following we use the cubic notations with $a_{p}$=3.84 {\AA} and hence the reciprocal-lattice vectors are given in
reduced lattice units as {\bf Q}=(h,k,l)=$\frac{2\pi}{a_{p}}$(q$_{x}$,q$_{y}$,q$_{z}$). {\ndsr} with x=0.49 and x=0.50 is a
paramagnetic-insulator at room-temperature and undergoes a transition into a FMM ordered state below {\tc}$\approx$250 K. Upon cooling a
transition of first order from the FMM state into a correlated charge, orbital and antiferromagnetic spin ordered state can be
observed.\cite{Kuwahara+95,Kawano+97,Kajimoto+99}

All of our neutron scattering experiments have been performed at the Laboratoire L\'eon Brillouin in Saclay. Diffraction experiments were made
using the 3T.1 diffractometer and the G4.3 spectrometer. Inelastic neutron data were collected with the 1T and 4F.2 spectrometers. The
spectrometer 1T, installed on the thermal channel of the reactor, was used with a pyrolytic graphite (PG) monochromator and with a PG analyzer.
To suppress higher harmonics two PG filters were installed and the final neutron energy was fixed at either \ef=14.7 meV or \ef=8.04 meV. The
cold three-axis spectrometer 4F.2 working with a PG double-monochromator was used with a Be-filter to suppress higher harmonics.

\begin{figure*}
\includegraphics[width=0.9\textwidth]{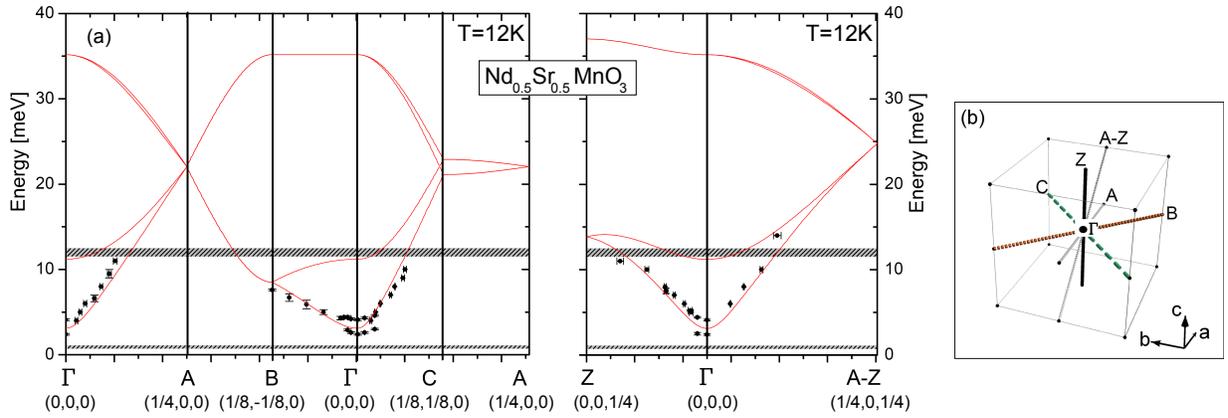}
\caption{(color online) Dispersion of the magnetic excitations in {\ndsrhalb} along the main-symmetry directions (a). The path $\Gamma$-A stands
for investigations in [1,0,0] direction (diagonal to the chains), while $\Gamma$-B measures the dispersion perpendicular and $\Gamma$-C parallel
to the zigzag chains. Investigations parallel to the c-axis are presented by $\Gamma$-Z. Solid curves are fits as described in the text and
crystal field excitations of the Nd ions are indicated by grey-shaded bars. The different paths along the main-symmetry directions around an
antiferromagnetic zone-center for chains propagation in [1,1,0] directions is shown in (b).} \label{full-ce}
\end{figure*}

\section{The charge, orbital and spin ordered state in N\MakeLowercase{d}$_{0.5}$S\MakeLowercase{r}$_{0.5}$M\MakeLowercase{n}O$_{3}$}

The ferromagnetic metallic and antiferromagnetic insulating states in our single-crystal of {\ndsrhalb} have been well characterized by
macroscopic techniques and by diffraction experiments demonstrating the high quality of our sample. Measurements of electric resistivity as a
function of temperature indicate a transition into a (half-)metallic state below {\tc}= 255 K and simultaneously a rapid increase of
magnetization points to ferromagnetic ordering [Fig. \ref{makr}(a),(d)]. The transition into a charge, orbital and spin ordered state at
{\tco}$\approx$160 K is of the first order. Herby, the electrical resistivity exhibits a jump by more than three orders of magnetiude and the
magnetization vanishes. X-ray powder diffraction has been performed from T=15 K up to T=850 K. The refinement of the data was carried out with
the Rietveld method implemented in the FullProf suite.\cite{Carvajal+93} The best description of the data can be obtained with the space group
Ibmm for the studied temperature range. The symmetry reduction from the space group Pm$\overline{3}$m of an ideal undistorted perovskite to
space group Ibmm arises from tilting of {MnO$_{6}$} octahedra around an axis parallel to the edges in the ab-planes. The orthorhombic distortion
of the lattice parameters increases rapidly at T=550 K which can arise from an orbital ordering effect of {\mnd} ions [Fig. \ref{makr}(c)]. Upon
cooling through the FMM phase transition, a slight increase of the orthorhombic c-axis with a shrinking of the a and b-axis is observed. It
appears likely that upon entering the FMM state an enhancement of the in-plane metallicity an FM exchange is favored. An abrupt change of the
lattice parameters during the transition into the COSO state indicates a sudden and strong structural change [Fig. \ref{makr}(b)].
Temperature-dependent neutron diffraction experiments at the Bragg reflection {\bf Q$_{FM}$}=(1,0,0) prove ferromagnetic ordering between
{\tco}$<$T$<${\tc} which is in perfect agreement with our magnetization data [Fig. \ref{makr}(e)]. Note, that the FM signal at {\bf
Q$_{FM}$}=(1,0,0) is also visible above {\tc}, but it gets strongly suppressed in the COSO state. The temperature-dependent behavior of the
superstructure reflection intensity of {\bf Q$_{Mn^{3+}}$}=(0.25,0.25,0.5) provides information on magnetic ordering of {\mnd} below {\tco}
[Fig. \ref{makr}(e); for more details see Ref. {\onlinecite{Senff+08}]. The quality of our sample is evident from the observed sharp
transitions, the increase in resistivity by several orders of magnitude, and from the perfect agreements of {\tc} and {\tco} with published
data.\cite{Kuwahara+95,Kawano+97} Although charge and orbital ordering cannot be directly detected by neutron diffraction, charge and orbital
ordering leads to a characteristic distortions of the MnO$_{6}$-octahedra that are visible in neutron diffraction experiments due to the higher
sensitivity to oxygen.

The Goodenough model predicts an alternating charge ordering of {\mnd} and {\mnv} ions. Considering an initially ideal cubic perovskite, this
doubles the structural unit cell to ($\sqrt 2a_{p}\times\sqrt 2a_{p}\times a_{p}$) and causes superstructure reflections with propagation vector
{\bf k}$_{\text{co}}$=($\pm \frac{1}{2}$,$\pm \frac{1}{2}$,0). Crystallographic studies find far smaller disproportion than a complete
{\mnd}$/${\mnv} (see Ref. \onlinecite{Jirak+00,Herrero+04}). Nevertheless we use the nominal labels {\mnd} and {\mnv} throughout this paper. The
orbital ordering of {\eg}-electrons on {\mnd}-sites reduces the symmetry to orthorhombic with lattice parameters $2\sqrt 2 a_{p}$ along [1,1,0]
and $\sqrt 2a_{p}$ along [-1,1,0] for zigzag chains running along [1,1,0]. As the zigzag chains can propagate along [1,1,0] and [-1,1,0],
orbital superstructure reflections with propagation vector {\bf k}$_{\text{oo}}$=($\pm \frac{1}{4}$,$\pm \frac{1}{4}$,0) can be detected. The
orthorhombic distortion induces a twofold twinning since the orbitals can stack either along [1,1,0] or [-1,1,0] direction. Both arrangements
superimpose and contribute to an equal amount in the scattering geometry studied in our sample. The superposition with the structural twinning
in these pseudo-cubic compounds can lead to a very complex twinned structure in the COSO state. The charge and orbital ordering implies an
antiferromagnetic CE-type ordering with ferromagnetic alignment of magnetic moments along the chains and antiferromagnetic coupling of adjacent
chains. Fig. \ref{ce-pattern} shows a schematic representation of the charge, orbital and spin ordering with zigzag chains propagation in
[1,1,0] direction. In perovskite manganites the planes of the orbital ordering always align with the structural distortions due to the
octahedron tilting. The latter leads to an unit cell $\sqrt{2}a_{p}\times\sqrt{2}a_{p}\times 2a_{p}$ with the doubling being always
perpendicular to the layers of orbital order (note that the zigzag chains run in these). Due to the strong structural distortion, orbitals and
charges align in rows along this perpendicular direction (c-axis). Charge and orbital order are thus not modulated along this direction,
corresponding to a ferro-type stacking. In the following the stacking direction is taken always along [0,0,1] in our cubic notation.\cite{Imma}
The magnetic CE-type spin ordering results in two sublattices: the magnetic unit cell of {\mnv} and the magnetic sublattice of {\mnd}. The
latter one is depicted in Fig. \ref{ce-pattern}. Superstructure reflections referring to the spin ordering of {\mnv} can be detected at
propagation vectors {\bf k}$_{\text{Mn}^{4+}}$=$\pm$($\frac{1}{2}$,0,$\frac{1}{2}$) or {\bf
k}$_{\text{Mn}^{4+}}$=$\pm$(0,$\frac{1}{2}$,$\frac{1}{2}$) and superstructure reflections referring to magnetic ordering of {\mnd} at {\bf
k}$_{\text{Mn}^{3+}}$=$\pm$($\frac{1}{4}$,$\pm \frac{1}{4}$,$\frac{1}{2}$). To investigate pure charge or orbital ordering we have to measure
superstructure reflections with large Q where magnetic order is negligible because the magnetic form-factor decreases rapidly with increasing
scattering vector.

To study the spin-wave dispersion in the AFM ordered state along the main-symmetry directions, we aligned the sample in two different
scattering-planes. Most of our experiments have been done by aligning the crystal in the (1,1,0)/(0,0,1) scattering plane to access momentum
transfers of the kind (h,h,l). In order to focus on the spin-wave dispersion parallel to the zigzag chains the sample was orientated in a less
conventional geometry with reflections (2,1,1)/(0,1,-1) defining the scattering plane. The spin-wave dispersion at T=12 K, observed by inelastic
neutron scattering from 2 meV up to about 12 meV is depicted in Fig. \ref{full-ce}. The solid lines correspond to a fit based on linear
spin-wave theory. Details on the theoretical calculations of the magnetic excitation spectrum are given below. Besides the superposition of all
twins rendering the study at high energies impossible, we observe Q-independent signals around 0.9 meV and around 12 meV pointing to crystalline
electric field (CEF) excitations of the Nd ions. Scans in different Brillouin zones and observations in similar compounds support this
conclusion.\cite{Petitgrand+97,Kirby+06,Krishnamurthy+06} Consequently clear magnetic excitations in all symmetry directions at 12 meV are not
recordable. Our spin-wave studies have been performed around several antiferromagnetic zone centers, $\Gamma =(\frac{h}{4},\frac{k}{4},\pm
\frac{1}{2})$ with h,k odd integer, and have been combined to the presented dispersion. Fig. \ref{scans} shows examples of typical raw-data
scans to study the spin-wave dispersion along the main-symmetry directions. We were able to separate between different paths since only one of
the orbital twins contributes to a quarter-indexed magnetic zone center. For example, to measure the spin-wave perpendicular to the zigzag
chains ($\Gamma$-B) starting at the zone center ($\frac{1}{4},\frac{3}{4},\frac{1}{2}$) one has to measure along the [-1,1,0] direction while
the [1,1,0] direction measures the path parallel to the zigzag chains ($\Gamma$-C) [Fig. \ref{full-ce}(b)]. Starting from the antiferromagnetic
zone center, we find a gap and a splitting into two single modes, similar to observations in the single-layered system {\lsmh} (Ref.
\onlinecite{Senff+06}). The degeneracy of the two AFM magnon branches seems to be removed due to the single-ion anisotropy. Energy scans at
several magnetic zone-centers exhibit two well distinguishable contributions at energies E$_{1}$=2.4(1) meV and E$_{2}$=4.1(1) meV [Fig.
\ref{scanszb}(a)]. The gap and energy splitting of the lowest excitations can be explained by the alignment of the magnetic moments. The
easy-axis of the magnetic moments of {\ndsrhalb} is in the ab-plane at an angle of 45$^{\circ}$ with respect to the a and b axis (cubic
notations).\cite{Kajimoto+99} To study the splitting near the zone-center and to follow the two modes in the directions $\Gamma$-B and
$\Gamma$-C we performed inelastic neutron experiments with the cold-source spectrometer. Away from the zone-center, at q$>$0.025, the two
excitations merge into a single signal.

\begin{figure}
\centering
\includegraphics[width=8.6cm]{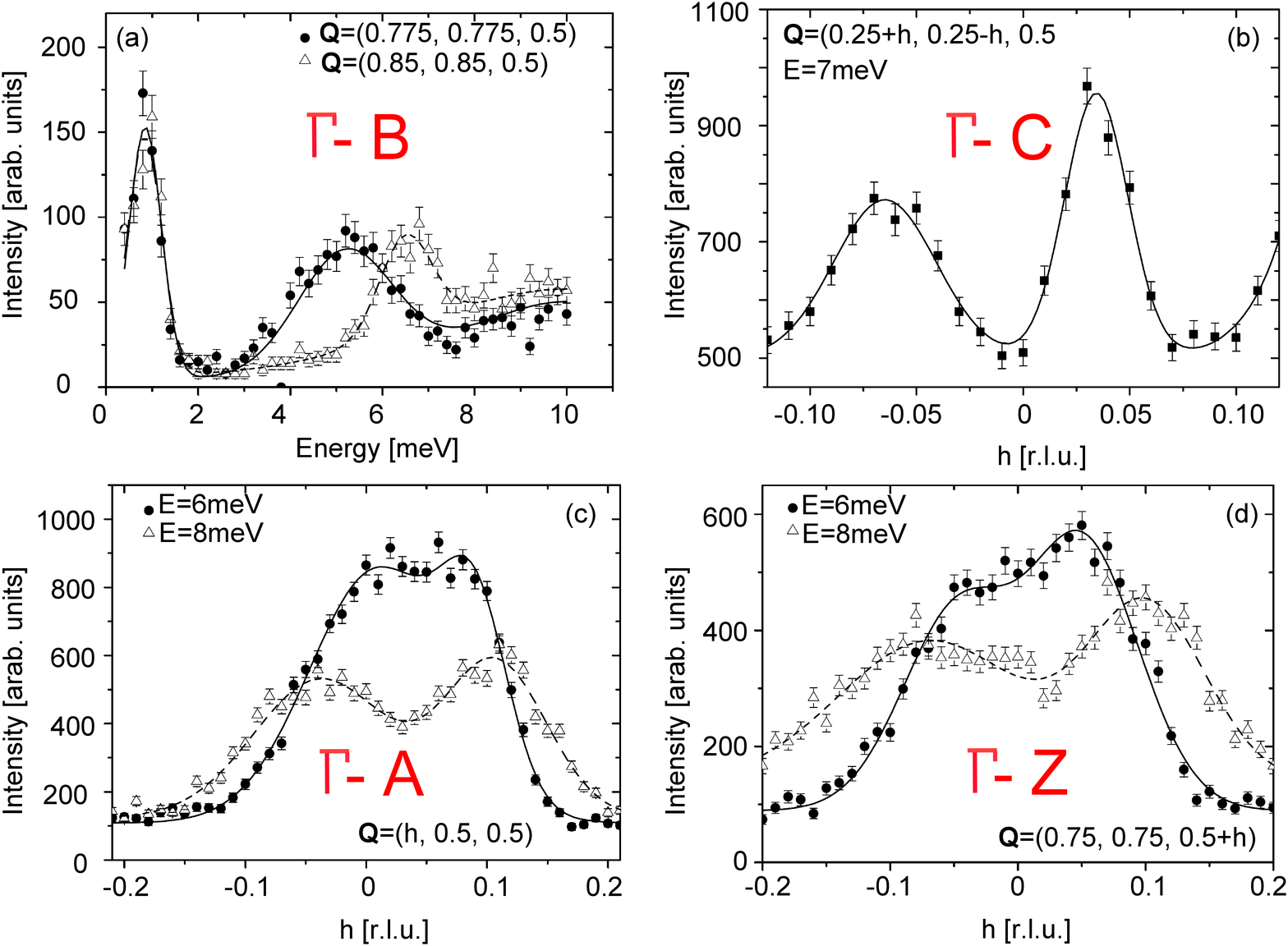}
\caption{(color online) Representative energy and constant-energy scans at T=12 K in {\ndsrhalb} along the main-symmetry directions. Two
constant-{\bf Q} scans which measures the dispersion along $\Gamma$-B are depicted in (a). Constant-energy scans to study the acoustic branch
parallel to the chains ($\Gamma$-C), 45$^{\circ}$ diagonal to the chains ($\Gamma$-A) and in c-direction ($\Gamma$-Z) are shown in (b)-(c).
Solid lines correspond to fits with Gaussians.} \label{scans}
\end{figure}

The lowest acoustic branch along the path $\Gamma$-C exhibits a steeper dispersion than that along the path $\Gamma$-B. The endpoint of the
acoustic branch at B can be observed at 7.5(1) meV. At this energy, we find the acoustic branch along $\Gamma$-C in the middle of the Brillouin
zone pointing to a stronger ferromagnetic coupling parallel to the zigzag chains in comparison to the coupling perpendicular to the chains.
Although magnetic excitations at the zone-boundary parallel to the chains can experimentally not be resolved due to the twinning, the
anisotropic behavior is apparent. Our theoretical calculations predict a magnon energy of about 22 meV at the zone boundary which is three times
larger than the endpoint B of the branch $\Gamma$-B. This study is in full agreement with the observations in {\lsmh} (Ref.
\onlinecite{Senff+06}). The antiferromagnetic coupling along the c-direction ($\Gamma$-Z) exhibits an intermediate dispersion: this means a
steeper acoustic branch in comparison to the antiferromagnetic coupling in $\Gamma$-B, but flatten than the ferromagnetic coupling along the
zigzag chains ($\Gamma$-C). Consequently the antiferromagnetic coupling between adjacent chains is weaker in comparison to the antiferromagnetic
stacking in c-direction. This is fundamentally different in the layered materials, where there is no significant c-axis coupling. The CE-type
order in {\ndsrhalb} is thus - as expected - much more three-dimensional in nature, the AFM stacking in c-direction is stronger than the AFM
coupling between zigzag chains in the ab-layers.

Theoretically, the formation of the CE-type superstructure close to half filling is understood in terms of the degenerate double-exchange (DDEX)
model.\cite{Brink+99,Feiner+99,Hotta+99,Efremov+04} The t$_{2g}$ electrons form local magnetic moments (core spins) which are coupled to the
itinerant electrons of the degenerate {\eg} orbitals by a ferromagnetic interatomic Hund's exchange $J_H$. In the context of the manganites, the
DDEX model can be greatly simplified since $J_H$ is much larger than the electronic bandwidth and the super-exchange between core spins which
are treated as classical. In the limit $J_H\to\infty$, the itinerant-electron spins are perfectly aligned with the local-moment spins.
Therefore, the itinerant electrons can move only in a ferromagnetic background. The complex CE superstructure is the result of the frustration
between the electronic kinetic energy and the antiferromagnetic super-exchange between the core spins. In this phase, the electronic kinetic
energy is directed along ferromagnetic zigzag chains which form because of a cooperative ordering of charge- and orbital degrees of
freedom.\cite{Senff+06}

\begin{figure}
\centering
\includegraphics[width=8.6cm]{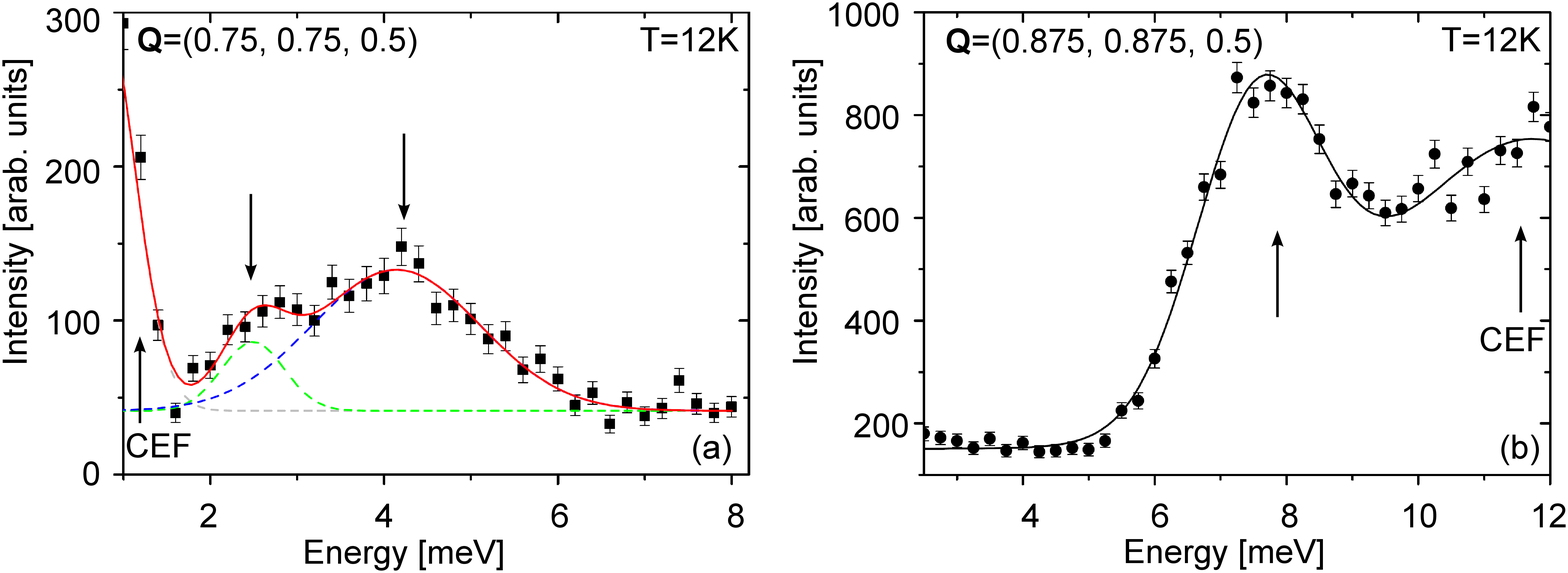}
\caption{(color online) Energy scans in the first Brillouin zone at the antiferromagnetic center {\bf Q}=(0.75,0.75,0.5) (a) and at the zone
boundary B along the acoustic branch $\Gamma$-B (b). The gap and the splitting of the excitations at the zone-center is clearly observable
(indicated by black arrows). The signal around E=1 meV corresponds to CEF excitations of the Nd ions. (The elastic line is not shown). The
excitation at the end-point B can be observed at 7.5(1) meV while the signal around 12 meV can be ascribed to a CEF, as well (b). Dashed lines
correspond to individual fits with Gaussians and the solid line to the resulting fit.} \label{scanszb}
\end{figure}

Although, the DDEX model treated in the semiclassical approximation can explain the richness of the phase diagram\cite{Efremov+04} and in
particular account for the CE structure observed in the close vicinity of half filling,\cite{Sternlieb+96,Radaelli+97,Kawano+97,Kajimoto+99} the
calculation of the magnetic excitation spectra within the DDEX model is extremely difficult since it requires finite $J_{H}$ and quantum
mechanical treatment of the core spins.\cite{Shannon+02,Lv+10} To avoid these complications, we compute the  spectrum from a spin-only model
with effective exchange couplings which includes both, AFM-super- and FM-double-exchange contributions. The latter are expected to contribute
predominantly along the chains and to overcompensate the bare AFM super-exchange. Consequently, a minimal effective model consists of FM
couplings $J_{FM}$ along the chains and AFM couplings $J_{AFM}$ between the chains.

Indeed, the magnetic excitation spectrum of the layered system {\lsmh} can be explained remarkably well by this minimal model.\cite{Senff+06}
Including an additional next-nearest neighbor exchange $J_{FM,2}$ between {\mnv} sites along the chains (Fig. \ref{fig:spinmodel}) and a small
single-ion anisotropy $\Lambda$ to account for the spin gap, one finds an excellent description of the neutron scattering data even for optical
branches at higher energies which has been interpreted as further evidence for the existence of the AFM CE-type superstructure at half filling.

In the following, we use exactly the same model to fit the spectrum of the perovskite {\ndsrhalb} including an additional coupling $J_{{AFM},c}$
along the $c$ direction (Fig. \ref{fig:spinmodel}). The effective spin-only Hamiltonian reads

\begin{eqnarray}
\hat{H} & = & -J_{FM} \sum_{\langle i,j\rangle_\parallel}\hat{\bf{S}}_i\hat{\bf{S}}_j -J_{FM,2} \sum_{\langle\langle
i,j\rangle\rangle_\parallel}^{\textrm{Mn}^{4+}}\hat{\bf{S}}_i\hat{\bf{S}}_j \nonumber \\
& + & J_{AFM}\sum_{\langle i,j\rangle_\perp}\hat{\bf{S}}_i\hat{\bf{S}}_j\nonumber +J_{{AFM},c}\sum_{\langle i,j\rangle_c}{\bf{S}}_i{\bf{S}}_j
\nonumber \\
& + &\Lambda\sum_i(\hat{S}^z_i)^2
\end{eqnarray}

where the exchange couplings $J_{FM}$, $J_{FM,2}$, $J_{AFM}$, $J_{{AFM},c}$ and single-ion anisotropy $\Lambda$ are defined to be positive. The
sites on the three dimensional lattice are labelled by $i$, $\langle\cdot,\cdot\rangle_\parallel$ and
$\langle\langle\cdot,\cdot\rangle\rangle_\parallel$ denote nearest and next-nearest neighbor bonds along the FM zigzag chains. Note that
$J_{FM,2}$ acts only between {\mnv} spins on the corners of the chains. Nearest-neighbor bonds between the chains in the $ab$-plane and along
the $c$ direction are denoted by $\langle\cdot,\cdot\rangle_\perp$ and $\langle\cdot,\cdot\rangle_c$, respectively. In order to facilitate the
comparison with Ref. \onlinecite{Senff+06} we use the same definition of the sum: each bond appears only ones.

In the following, we use linear spin-wave theory to calculate the magnetic excitation spectrum and subsequently fit the experimental data. The
procedure is standard and details can be found elsewhere.\cite{Ventura+03,Sikora+05} After representing the spins by Holstein-Primakoff bosons
the spectrum is obtained by a Bogoliubov transformation which amounts to diagonalizing a 16 by 16 matrix which is explicitly given in Ref.
[\onlinecite{Ventura+03}]  in the case $J_{FM,2}=\Lambda=0$.

Whereas the idealized CE structure assumes a checkerboard ordering of  {\mnd} and {\mnv} sites with corresponding local moment spins of $S_1=2$
and $S_2=3/2$, respectively, the experimentally observed charge and spin modulations are significantly smaller. To account for this, in the
following we use $S_1=1.79$ and $S_2=1.67$ such that $S_1S_2=3$ and $S_1/S_2=3/2.8$ equal to the ratio of the measured
moments.\cite{Kawano+97,Kajimoto+99}

The best fit parameters are summarized in Table \ref{afmtable}. As in the layered system,\cite{Senff+06} the FM coupling along the chains
$J_{FM}$ is found to be much larger than the AFM in-plane coupling $J_{AFM}$ between the chains. For the {\lsmh} compound we used for the spins
of {\mnd} $S_1=2$ and for the spins of {\mnv} $S_2=1.5$. As the product of $S_1S_2=3$ is the same for both systems, the nearest-neighbor
parameters $J_{FM}$ and $J_{AFM}$ can directly be compared.

This observation and the ratio $J_{{AFM},c}/J_{AFM}\approx 4$ are in agreement with a quantitative estimate of the exchange couplings based on a
modified Kanomori model which carefully takes details of the lattice structures of various CE-type manganites into account.\cite{Gontchar+02}

\begin{table}[h]
  \centering
    \begin{tabular}{c|c|c}

\multicolumn{1}{c}{} & \multicolumn{1}{|c|}{\ndsrhalb} & \multicolumn{1}{c}{\lsmh}\\

\hline
$J_{FM}$ [meV]            &      6.76          &           9.98   \\
$J_{FM,2}$ [meV]           &      2.76         &            1.83   \\
$J_{AFM}$ [meV]           &      0.75         &            3.69   \\
$J_{AFM,c}$ [meV]         &      3.22         &            --  \\

\end{tabular}
  \caption{Results of the obtained fit parameters in the AFM CE-type state for {\ndsrhalb} in comparison to the fit parameters in {\lsmh} from
  Ref. \onlinecite{Senff+06}.
  } \label{afmtable}
\end{table}

\begin{figure}
\centering
\includegraphics[width=5cm]{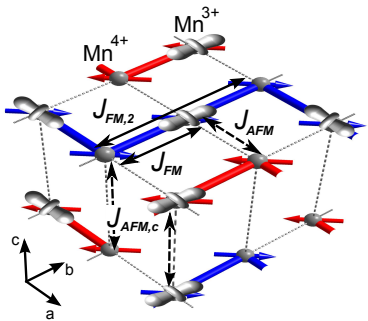}
\caption{(color online) Illustration of the effective spin-only model for the AFM CE-type structure. Spins along the zigzag chains are coupled
by ferromagnetic nearest and next-nearest neighbor exchanges $J_{FM}$ and $J_{FM,2}$, the latter acting only between {\mnv} sites. The chains
are coupled antiferromagnetically by exchanges $J_{AFM}$ and $J_{{AFM},c}$ between neighbors in the $ab$-plane and along the $c$-axis,
respectively.} \label{fig:spinmodel}
\end{figure}

$J_{{AFM},c}$ connects either two {\mnd} or two {\mnv} sites, whereas $J_{AFM}$ couples one {\mnd} and one {\mnv} site where imperfect orbital
polarization will imply a FM double-exchange interaction rendering the AFM super-exchange. As in addition the crystal lattice is considerably
flattened, $J_{{AFM},c}$ is much larger than $J_{AFM}$. In our model we include a simple single-ion anisotropy term
$\Lambda\sum_i(\hat{S}^z_i)^2$. The ordered moments in the CE-type phase in {\ndsrhalb} are aligned along the ab-plane. Therefore an anisotropy
term perfectly accounts for an ab-easy-plane system. However the spins seem to be well fixed also within the ab-layers causing a splitting in
the zone-center magnon energies.\cite{Kawano-F+03} The same situation is also found in single-layered {\lsmh}, where the lower excitation could
be assigned to the in-plane anisotropy. Also in {\ndsrhalb} we may identity the lower excitation with the in-plane anisotropy. It costs more
energy to rotate the spins into the c-direction. A full treatment for such complex anisotropy has been given for a simple AFM layered structure
in Ref. \onlinecite{Jongh+90}.

Inelastic neutron scattering experiment have been able to determine the spin-wave excitations of {\ndsrhalb} for energies up to about 12 meV.
Our data exhibit a steeper magnetic dispersion parallel to the zigzag chains while the dispersion perpendicular to the chains is flatten. This
observation is in contrast to the Zener-polaron model which implies a strong coupling inside the dimers. Taking this and the magnetic structure
of the Zener-polaron scenario into account one obtains just the opposite ansisotropy in the dispersion slopes along the $\Gamma$-B and
$\Gamma$-C directions studied here. The spin-wave theory within the CE-type structure explains perfectly our experimental data. The experimental
and the theoretical analysis are in agreement with the single-layered compound {\lsmh} (Ref. \onlinecite{Senff+06}). Consequently we may extend
the strong support of the Goodenough model to perovskite manganites around half-doping.

\section{The ferromagnetic metallic state in N\MakeLowercase{d}$_{0.50}$S\MakeLowercase{r}$_{0.50}$M\MakeLowercase{n}O$_{3}$}

\begin{figure}
\includegraphics[width=0.47\textwidth]{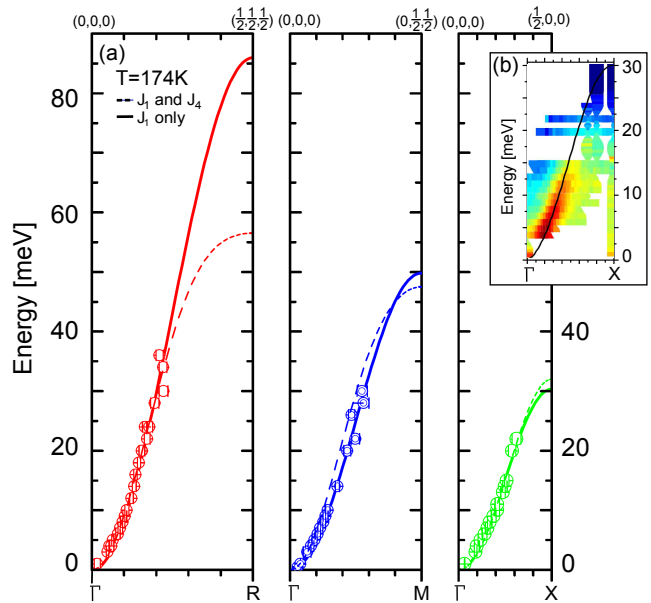}
\caption{(color online) Full spin-wave dispersion along the main-symmetry directions in the FMM state in {\ndsrhalb} at T=174 K. Experimental
data were taken along the directions $\Gamma$-R, $\Gamma$-M, and $\Gamma$-X, respectively. Solid lines refer to a fit with a Heisenberg
ferromagnet with only nearest-neighbor interactions and dotted lines to a fit with nearest and forth-nearest-neighbor interactions. A
significant improvement of the fit to the data by taking additionally $J_{4}$ into account can not be observed. An enhanced signal around the
zone boundary might be given by several magnetic levels (b). For more details see text.} \label{full-fm-05}
\end{figure}

In order to analyze the competition between the insulating COSO and the FMM phases in {\ndsrhalb} we performed studies of the magnon dispersion
in the FMM phase as well. The magnetic excitations in the FMM state in perovskite manganites have been studied in many different compounds and a
damping of the spin-wave dispersion towards the zone boundaries have been reported in many cases.
\cite{Perring+96,Hwang+98,Fernandez+98,Endoh+05,Ye+06,Ye+07,Petit+09} To explain this behavior, several microscopic mechanisms have been
proposed, without reaching a clear conclusion (for a brief summary see Ref. \onlinecite{Zhang+07}). One explanation for the softening of the
spin-wave dispersion might be the development of quantized spin-waves within ferromagnetic clusters of restricted sizes as reported in Ref.
\onlinecite{Hennion+05,Petit+09}.

We studied the spin dynamics of {\ndsrhalb} at T=174 K, where the ferromagnetic signal of the reflection at {\bf Q$_{FM}$}=(1,0,0) exhibits its
maximum [see Fig. \ref{makr}(e)]. To investigate magnetic excitations along the main-symmetry directions [1,0,0], [0,1,1] and [1,1,1], we
aligned the sample with reflections (1,0,0) and (0,1,1) defining the scattering plane. All data have been collected around the Bragg-reflections
{\bf Q$_{FM}$}=(1,0,0), {\bf Q$_{FM}$}=(0,1,1) and {\bf Q$_{FM}$}=(1,1,1). Most data have been collected with the thermal spectrometer 1T with
fixed E$_{f}$=14.7 meV. The calibration of the spectrometer has been checked by a Vanadium standard crystal. An energy scan with E$_{f}$=14.7
meV indicates a resolution of $\Delta$E$\approx$1 meV. Fig. \ref{full-fm-05} presents the spin-wave dispersion along the main symmetry
directions. The data have been corrected by the Bose factor and finally fitted with the ResLib code which calculates the convolution of the
four-dimensional resolution function with the dispersion.\cite{Reslib} We were able to investigate magnetic excitations of the acoustic branch
along the pathes $\Gamma$-R, $\Gamma$-M, and $\Gamma$-X. An energy scan with the cold-source spectrometer 4F.2 at the zone-center {\bf
Q$_{FM}$}=(1,0,0) yields no evidence for an excitation gap. Constant-energy scans around the ferromagnetic zone-center {\bf Q$_{FM}$}=(1,0,0)
are depicted in Fig. \ref{mag-phon-fm05}(a). Investigations in different Brillouin zones clearly identify a magnetic  origin for the signals
appearing at low energy transfer, since the intensity follows the magnetic form factor as shown in Fig. \ref{mag-phon-fm05}(b). In Ref.
\onlinecite{Kawano-F+03} the spin-wave dispersion in the FMM state in {\ndsrhalb} along different symmetry directions has been studied and an
extremely flat dispersion with energies around 10 meV at the zone boundaries is reported. Our experiments disagree with theses observations
since our data yield much steeper branches. For example, clear excitations along the [1,1,1] direction between energies of E=18 meV and E=22 meV
can be observed in the middle of the Brillouin zone (see Fig. \ref{hscans-fm05}). The main magnetic signal seems to stem from the simple
ferromagnetic magnon branch; but there are weak but significant additional signals which are most likely magnetic as well. The constant energy
cuts (h,0,0) across the (1,0,0) FM Bragg-peak show more intensity near the magnetic zone boundaries $\approx$(0.5,0,0) than at the center which
can not be attributed to the background. Also the energy scan at (1.5,0,0) exhibits significant intensity below the energy of the magnon branch
[Fig. \ref{full-fm-05} (b)]. This scattering resembles the splitting of the magnon branches observed in La$_{1-x}$(Ca,Sr)$_{x}$MnO$_{3}$ at
lower doping, which was attributed to a confinement.\cite{Hennion+05,Petit+09} However, such confinement effect - if present - must be much
weaker in this higher-doped {\ndsrhalb} material.

Spin-wave dispersion in 113-manganites with a large electron bandwidth and high {\tc} can be described by calculations with the conventional
Heisenberg model including only the nearest-neighbor exchange interactions.\cite{Martin+96,Perring+96} In the following the description of our
data takes only the sharp signals into account without keeping the increasing intensity near the zone-boundaries in mind. To evaluate the
spin-wave dispersion in {\ndsrhalb} we use two models: (1) we consider only the ferromagnetic coupling between nearest-neighbor interaction
$J_{1}$, (2) nearest-neighbor interactions $J_{1}$ and forth-nearest interactions $J_{4}$ are taken into account.

\begin{figure}
\includegraphics[width=0.47\textwidth]{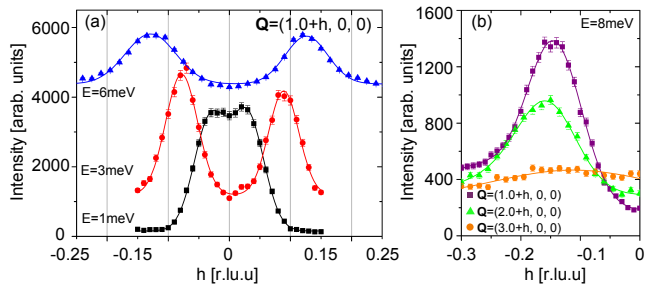}
\caption{(color online) Constant-energy scans around the zone-center {\bf Q$_{FM}$}=(1,0,0) in {\ndsrhalb} along the symmetry direction [1,0,0].
Data at E=1 meV and 3 meV are recorded with the spectrometer 1T with fixed \ef=8.05 meV and at E=8 meV with fixed \ef=14.7 meV, respectively
(a). A decrease of the signal with increasing scattering vector demonstrates its magnetically origin (b).} \label{mag-phon-fm05}
\end{figure}

The Hamiltonian of a Heisenberg ferromagnet can be written as:

\begin{equation}
\label{hb1} \hat{H} \,=\,-\sum_{ij} J_{ij} {\bf S}_{i} {\bf S}_{j}
\end{equation}
Herby $J_{ij}$ describes the magnetic exchange coupling between two spins ${\bf S}_{i}$ and ${\bf S}_{j}$ at the position ${\bf R}_{i}$ and
${\bf R}_{j}$. Note that a pair appears twice in the sum. \cite{pair} In linear approximation the spin-wave dispersion with only ferromagnetic
coupling ($J_{ij}>0$) becomes:

\begin{equation}
\label{hb2} E({\bf q})\,=\, \hbar \omega({\bf q})=\,\Delta\,+\,2\,S\,[J({\bf 0})-J({\bf q}) ]
\end{equation}

with:

\begin{equation*}
\label{hb3}
 J({\bf q})\, = \, \sum_{j}J_{ij} e^{i{\bf q}({\bf R}_{i}-{\bf R}_{j})}
\end{equation*}

A finite energy gap near the zone-center can be controlled by $\Delta$. Since we do not observe a finite gap for the [1,0,0], [0,1,1] and
[1,1,1] directions, we will set $\Delta$ equal to zero.

The simplest case with only nearest-neighbor $J_{1}$ interactions leads to the relation:

\begin{equation}
\label{1neighb}
 E({\bf q})\, = \, \Delta+4SJ_{1}[3-\cos(q_{h}a)-\cos(q_{k}a)-\cos(q_{l}a)]
\end{equation}

Taking only $J_{1}$ into account, we can very well describe our experimental data (Fig. \ref{full-fm-05} solid lines). The fits of the data with
only the nearest-neighbor interaction $J_{1}$ yield $2SJ_{1}$=7.6(2) meV for the [1,0,0], $2SJ_{1}$=6.3(2) meV for the [0,1,1] and
$2SJ_{1}$=7.2(2) meV for the [1,1,1] direction. To compare an improvement of the description of the fit by introducing the coupling $J_{4}$ as
suggested previously in Ref. \onlinecite{Hwang+98}, the dispersion for the [1,0,0] direction takes the form [Fig. \ref{full-fm-049}(b)]:

\begin{eqnarray}
 E({\bf q})\, = \,\Delta\,+\,4SJ_{1}[(1-\cos(2 \pi q_{h}))]\,+ \nonumber \\
 4SJ_{4}[(1-\cos(4\pi q_{h}))]
 \label{4neighb}
\end{eqnarray}

\begin{figure}
\includegraphics[width=0.36\textwidth]{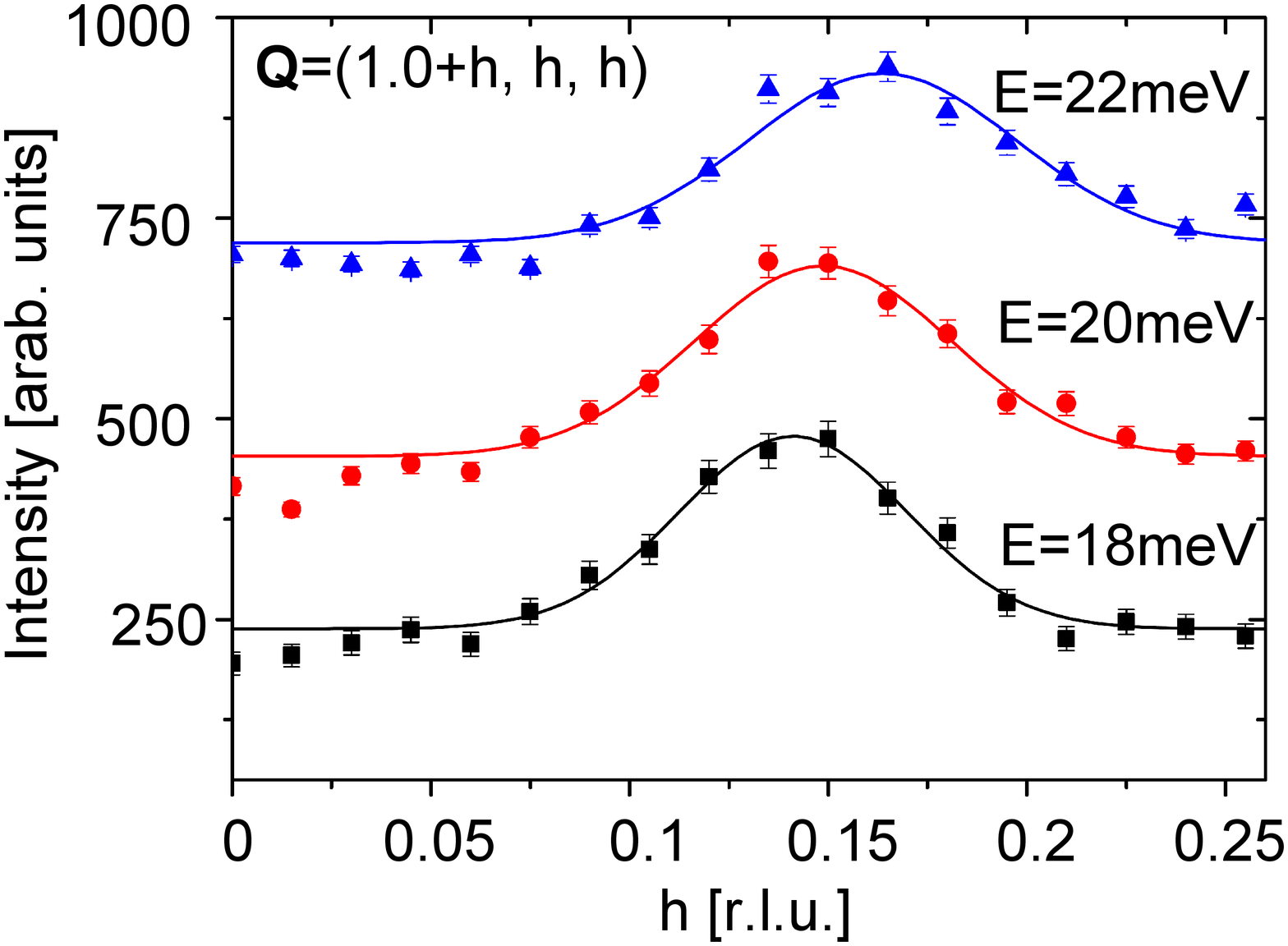}
\caption{(color online) Observed excitation at high energies along the symmetry direction [1,1,1]. The maximum shifts to increasing q with
increasing energy indicating the steeper dispersion.} \label{hscans-fm05}
\end{figure}

The best fit to the data with nearest-neighbor and forth-nearest interaction interaction is depicted in Fig. \ref{full-fm-05} by the dotted
lines. There is only little difference in the spin-wave dispersion of the two models in the energy range studied. Therefore our data give no
support for a related $2SJ_{4}$ parameter.

In the CE-type state we obtained the best fit to the data for the ferromagnetic coupling parallel to the chains with $2SJ_{FM}$=11.7 meV. In the
FM state we can describe the data with a mean exchange energy of $2SJ_{1}$=7.0(5) meV for the fit with nearest-neighbor interaction taking into
account. This means the nearest-neighbor ferromagnetic exchange in the FM state is strongly reduced by a factor of about 1.5 in comparison to
the AFM CE-type state, but note that each spin has six ferromagnetically coupled spins in the FMM state, compared to only two FM neighbors in
the CE-phases. In the single-layered compounds {\lsmh} an exchange energy of $2SJ_{FM}$=7.5(5) meV has been observed above the AFM CE-type
ordered state.\cite{Senff+08} Consequently, the exchange parameters in the FM states in {\lsmh} and in {\ndsrhalb} are almost identical. This
result is quite interesting since the single-layered compound does not show metallic behavior and possesses less neighbors. More importantly,
the layered compound only exhibits ferromagnetic fluctuations generated by ferromagnetic clusters with finite size above {\tco}. Our
investigation in the FMM state in {\ndsrhalb} indicate a quite good description of the data along the main-symmetry directions by taking only
the nearest-neighbor coupling $J_{1}$ into account (but note that our model is slightly over-simplified since we do not take the intensities
near the zone-boundary taking into account).

To get a more precise perception of the magnetic excitation in the FMM state in {\ndsr} and to compare the exchange parameters we decided to
study the spin-wave dispersion in a slightly electron-doped single crystal with a composition of {\ndsrvier}.

\begin{figure}
\includegraphics[width=0.54\textwidth]{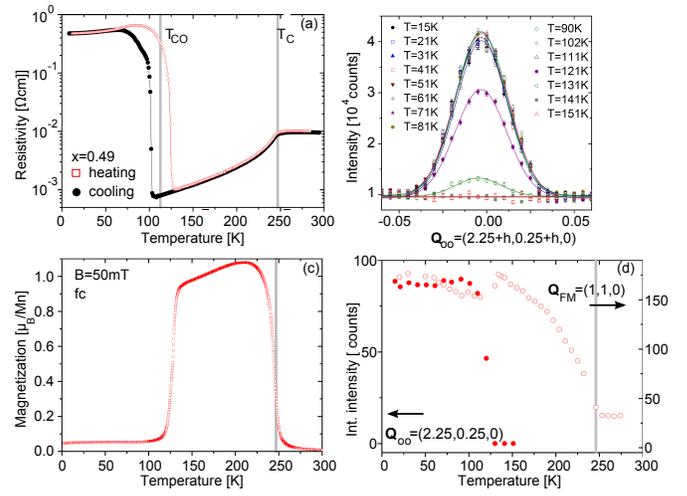}
\caption{(color online) Characterization of the single crystal {\ndsrvier} by macroscopic and microscopic studies. A ferromagnetic metallic
state can be observed by electric resistivity measurements and magnetization data (a,c). Elastic neutron scattering at {\bf Q$_{FM}$}=(1,1,0)
confirms ferromagnetic ordering below {\tc} (d). A sudden increase in intensity of the superstructure reflection {\bf Q$_{OO}$}=(2.25,0.25,0)
below {\tco} refers to orbital ordering (b),(d).} \label{makr049}
\end{figure}

\section{The ferromagnetic metallic state in N\MakeLowercase{d}$_{0.51}$S\MakeLowercase{r}$_{0.49}$M\MakeLowercase{n}O$_{3}$}

\begin{figure}
\includegraphics[width=0.48\textwidth]{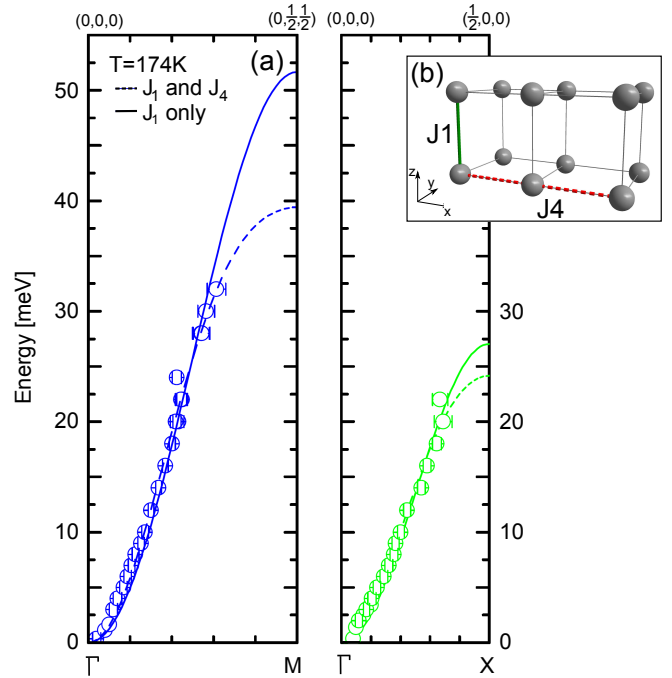}
\caption{(color online) Dispersion of the magnon excitation in the FMM state in {\ndsrvier} along the symmetry directions [1,0,0] and [0,1,1] at
T=151 K (a). Solid lines refer to fits with only nearest-neighbor interactions $J_{1}$ and dashed lines point to calculations with nearest and
forth-nearest-neighbor interactions $J_{1}$ and $J_{4}$ as depicted in (b).} \label{full-fm-049}
\end{figure}

A single-crystal of {\ndsr} with doping level x=0.49 has been characterized by temperature-dependent measurements of magnetization, resistivity,
X-ray, neutron diffraction and by EDX analysis. A significant drop of resistivity accompanied by an increase of magnetization reminiscent of the
FMM state is observed at {\tc}$\approx$245 K (Fig. \ref{makr049}). In accordance with the macroscopic studies, the scattering at {\bf
Q$_{FM}$}=(1,1,0) exhibits an increase of intensity below {\tc} which can be ascribed to ferromagnetic correlations [Fig. \ref{makr049}(c)].
Upon further cooling, the electric resistivity increases by more than two orders of magnitude at the first order transition into the COSO state
at {\tco}$\approx$112 K. Temperature-dependent investigations of the superstructure reflection referring to orbital ordering {\bf
Q$_{OO}$}=(2.25,0.25,0) yield a sizeable signal below {\tco} [Fig. \ref{makr049}(d)]. Comparing the transition temperatures into the COSO state
in {\ndsr} for x=0.50 and x=0.49, we observe a less robust COSO state in {\ndsrvier}. Before discussing inelastic data we want to compare the
ground state properties of the ideally doped system (x=0.5) in comparison to the slightly electron-doped system (x=0.49). In {\ndsrvier} below
T$\approx$50 K, the electric resistivity is almost constant. In contrast, for x=0.5 we find a continuous increase of resistivity below {\tco} up
to $\rho{\approx}10^{4}$ $\Omega$m at the lowest temperature in comparison to $\rho{\approx}10^{0}$ $\Omega$m for x=0.49. The slight doping
difference between these two compounds strongly affects the magnetization at low temperatures as well. The magnetization with an applied field
of B=50 mT saturates with m=0.05 $\mu_{ B}$/Mn for x=0.49 while the magnetization for x=0.50 tends to zero at low temperatures. These properties
underline the coexistence of antiferromagnetic CE-type and ferromagnetic ordering in the ground state of {\ndsrvier} as reported in Ref.
\onlinecite{Kajimoto+99}. Neutron diffraction of the reflection {\bf Q$_{FM}$}=(1,1,0) further supports this interpretation. The signal below
{\tc} increases as in the half-doped system, but in the COSO state in x=0.49 a high intensity can still be recorded at low temperature. We
analyzed the stoichiometry of the samples with EDX analysis. However the accuracy of this method does not allow for determining the content of
oxygen concentration. Hence we may not rule out an oxygen-deficiency. Concerning the various transition temperatures and the temperature
dependence of the electrical resistivity our sample is fully comparable to the {\ndsrvier} crystal studied in Ref. \onlinecite{Kajimoto+99}.

\begin{figure}
\includegraphics[width=0.48\textwidth]{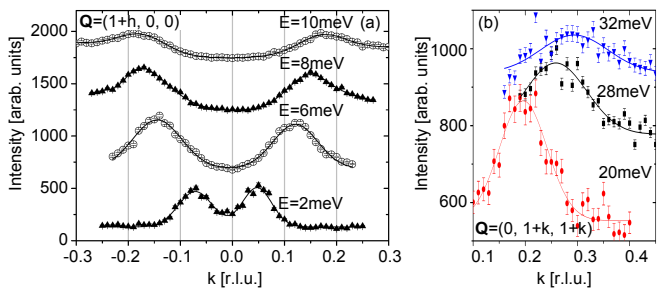}
\caption{Raw-data scans with fixed energy observed in [1,0,0] direction. Data at 2 meV are taken with {\ef}=8.05 meV and above with {\ef}=14.7
meV. Intensities are normalized to same monitor but shifted as guide to the eyes (a). Signals along the symmetry direction [0,1,1] exhibit a
loss in intensity and increase of width with increasing energy, too (b).} \label{raw049}
\end{figure}

We have studied the spin-wave excitations in {\ndsrvier} at T=151 K, which is well in the ferromagnetically ordered state, and present an
analysis of the data with the same Heisenberg models as those used for the half-doped compound. The scattering experiments have been performed
by aligning the crystal with the reflections (1,0,0)/(0,1,1), allowing momentum transfers of the kind (h,l,l). We focused our measurement on
excitations along the symmetry directions [1,0,0] and [0,1,1] in cubic notation. Data have been measured in the Brillouin zones around the
zone-centers {\bf Q$_{FM}$}=(1,0,0) and {\bf Q$_{FM}$}=(0,1,1) and were finally combined. The data have been corrected by the Bose factor and
finally fitted with the convolution of the four-dimensional resolution function with the dispersion. Fits to the data results in the spin-wave
dispersion as shown in Fig. \ref{full-fm-049}. Raw-data scans with fixed energy around ferromagnetic zone centers along the [1,0,0] and [0,1,1]
directions are depicted in Fig. \ref{raw049}.

In the following we will proof the experimental data of the sharp signals with the Heisenberg model as described for the half-doped system. The
investigations of the dispersion near the zone center $\Gamma$ with the cold triple-axis spectrometer are in agreement with our observations in
{\ndsrhalb} and we find no indication for a spin-wave gap in {\ndsrvier}. Consequently, we set $\Delta$ equal to zero in our calculation. To
evaluate the spin-wave dispersion in {\ndsrvier} we make the same assumptions as those used for the x=0.5 compound: (1) we consider only the
ferromagnetic coupling between nearest-neighbor interaction $J_{1}$, and (2) nearest-neighbor interactions $J_{1}$ and forth-nearest
interactions $J_{4}$ are taken into account.

All fit parameters for the spin-wave dispersion in {\ndsrvier} and {\ndsrhalb} are summarized in Table \ref{fmtable}. Since no significant
improvement of the reduced chi-squared ${\chi}_{red}^2$ for the model including $J_{4}$ and since the ratio of $J_{4}$/$J_{1}$ is quite small,
there is no evidence that the forth-nearest-neighbors is needed to describe the ferromagnetic spin-wave dispersion in {\ndsr} with x=0.49.

\begin{table}[h]
  \centering
    \begin{tabular}{c|c|c}

\multicolumn{1}{c}{ } & \multicolumn{1}{|c|}{{\ndsrvier}} & \multicolumn{1}{c}{{\ndsrhalb}}\\
              &[1,0,0] \hspace{0.35cm} [0,1,1]  & [1,0,0] \hspace{0.35cm} [0,1,1] \hspace{0.35cm} [1,1,1]   \\

\hline
model 1             & &  \\
$2SJ_{1}$ [meV] &6.8(3) \hspace{0.38cm} 6.5(3) & 7.6(2) \hspace{0.38cm} 6.3(2) \hspace{0.38cm} 7.2(2)   \\
\hline
model 2             &  &    \\
$2SJ_{1}$ [meV] &6.0(3) \hspace{0.38cm} 4.9(2) & 8.0(3) \hspace{0.38cm} 5.3(9) \hspace{0.38cm} 2.1(9)   \\
$2SJ_{4}$ [meV] &0.5(2) \hspace{0.38cm} 0.8(1) & -0.2(2) \hspace{0.38cm} 0.4(4) \hspace{0.38cm} 0.4(3)   \\
$J_{4}$/$J_{1}$ [\%] &8 \hspace{1cm} 16 & 2 \hspace{1cm} 8 \hspace{1cm} 19   \\

\end{tabular}
  \caption{Fit parameters for the spin-wave dispersion in {\ndsr} with x=0.49 and x=0.50 in the FMM state. The data have been obtained from two different
  models. Model 1, only nearest-neighbor interactions $J_{1}$ and model 2, nearest-neighbor interactions $J_{1}$ and forth-nearest interactions
  $J_{4}$ are taken into account.
  } \label{fmtable}
\end{table}

Our studies for x=0.49 yield the identical dispersion in comparison to the optimal-doped compound x=0.5 in the FMM state. Near the zone-center
the signal is sharp and resolution limited. One can easily notice that the intensity with increasing energy shrinks and the width starts to
increase. Following the sharp excitations with increasing energy, toward the zone boundary the signal is hampered, similar to observations in
the optimal doped compound x=0.50.

The observation of nearly identical exchange parameters in {\ndsr} with x=0.49 and x=0.50 again emphasizes an identical FMM state in x=0.49 and
x=0.50. The FMM state in {\ndsr} seems to be less sensitive to the doping than the COSO phase which is severely damaged by the small amount of
holes in {\ndsrvier}.

\section{Summary}

In conclusion, we have studied the microscopic origin of the COSO state in {\ndsrhalb} and the FMM state in {\ndsr} with x=0.50 and x=0.49 by
inelastic neutron scattering. Despite the complications arising from the twinning in these perovskite manganites with pseudo-cubic structure, we
were able to observe excitations along the main-symmetry directions up to E$\approx$12 meV in the AFM ground state of x=0.5. We clearly find an
anisotropic dispersion between the directions parallel and perpendicular to the orbital zigzag chains. Theoretical descriptions of the spin-wave
dispersion by the AFM CE-type ordering perfectly describe our experimental data yielding results similar to the single-layered compound {\lsmh}
(Ref. \onlinecite{Senff+06}). The good description of the experimental dispersion gives strong support for the qualitative CE-type Goodenough
model. A model with bond-centered charge order is even unable to explain the sign of the anisotropy. Although the Zener-polaron picture cannot
explain the data, a superposition of both, Goodenough model and Zener-polaron as suggested in Ref. \onlinecite{Efremov+04} can not fully be
excluded. The stacking of magnetic order along the c-direction, i.e. perpendicular to the planes of the zigzag chains, is rather strong. This
finding can be understood by the flattening of the crystal structure and by the fact this stacking arises from strong {\mnd}-{\mnd} and
{\mnv}-{\mnv} interactions.

The FMM state has been investigated in the half-doped compound {\ndsrhalb} and in the slightly electron-doped system {\ndsrvier}. Both
concentrations exhibit nearly the same spin-wave spectra, proving that the FMM state is little affected by the slight difference in doping. In
{\ndsrhalb} the magnon-dispersion along the main symmetry-directions [1,0,0], [0,1,1] and [1,1,1] can be observed. The FMM state seems to
consist of two parts; (1) a spin-wave dispersion which can be easily described with a Heisenberg model taking only nearest-neighbors
interactions into account and (2) weak additional intensities near the zone boundaries over a wide energy band. The spin-wave dispersion in
{\ndsrvier} along the [1,0,0] and [0,1,1] directions exhibits identical results with a steep dispersion and damped excitations near the
zone-boundary.

\subsection*{Acknowledgements}

This work was supported by the Deutsche Forschungsgemeinschaft through the Sonderforschungsbereich 608. We thank S. Heijligen for the SQUID
measurements. F. Kr\"uger acknowledges financial support from EPSRC under grant code EP/I004831/1. A. Nugroho thanks Cologne University and SFB
608 for hospitality.


\begin{thebibliography}{54}
\expandafter\ifx\csname natexlab\endcsname\relax\def\natexlab#1{#1}\fi \expandafter\ifx\csname bibnamefont\endcsname\relax
  \def\bibnamefont#1{#1}\fi
\expandafter\ifx\csname bibfnamefont\endcsname\relax
  \def\bibfnamefont#1{#1}\fi
\expandafter\ifx\csname citenamefont\endcsname\relax
  \def\citenamefont#1{#1}\fi
\expandafter\ifx\csname url\endcsname\relax
  \def\url#1{\texttt{#1}}\fi
\expandafter\ifx\csname urlprefix\endcsname\relax\def\urlprefix{URL }\fi \providecommand{\bibinfo}[2]{#2}
\providecommand{\eprint}[2][]{\url{#2}}

\bibitem[{\citenamefont{Jonker and Santen}(1950)}]{jonk+50}
\bibinfo{author}{\bibfnamefont{G.~H.} \bibnamefont{Jonker}} \bibnamefont{and}
  \bibinfo{author}{\bibfnamefont{J.~H.~V.} \bibnamefont{Santen}},
  \bibinfo{journal}{Physica} \textbf{\bibinfo{volume}{16}},
  \bibinfo{pages}{337} (\bibinfo{year}{1950}).

\bibitem[{\citenamefont{Santen and Jonker}(1950)}]{sant+50}
\bibinfo{author}{\bibfnamefont{J.~H.~V.} \bibnamefont{Santen}}
  \bibnamefont{and} \bibinfo{author}{\bibfnamefont{G.~H.}
  \bibnamefont{Jonker}}, \bibinfo{journal}{Physica}
  \textbf{\bibinfo{volume}{16}}, \bibinfo{pages}{599} (\bibinfo{year}{1950}).

\bibitem[{\citenamefont{Tokura and Tomioka}(1999)}]{Tokura+99}
\bibinfo{author}{\bibfnamefont{Y.}~\bibnamefont{Tokura}} \bibnamefont{and}
  \bibinfo{author}{\bibfnamefont{Y.}~\bibnamefont{Tomioka}},
  \bibinfo{journal}{J. Magn. Mater.} \textbf{\bibinfo{volume}{200}},
  \bibinfo{pages}{1} (\bibinfo{year}{1999}).

\bibitem[{\citenamefont{Radaelli et~al.}(1997)\citenamefont{Radaelli, Cox,
  Marezio, and Cheong}}]{Radaelli+97}
\bibinfo{author}{\bibfnamefont{P.~G.} \bibnamefont{Radaelli}},
  \bibinfo{author}{\bibfnamefont{D.~E.} \bibnamefont{Cox}},
  \bibinfo{author}{\bibfnamefont{M.}~\bibnamefont{Marezio}}, \bibnamefont{and}
  \bibinfo{author}{\bibfnamefont{S.-W.} \bibnamefont{Cheong}},
  \bibinfo{journal}{Phys. Rev. B} \textbf{\bibinfo{volume}{55}},
  \bibinfo{pages}{3015} (\bibinfo{year}{1997}).

\bibitem[{\citenamefont{Kawano et~al.}(1998)\citenamefont{Kawano, Kajimoto,
  Yoshizawa, Fernandez-Baca, Tomioka, Kuwahara, and Tokura}}]{kawa+98}
\bibinfo{author}{\bibfnamefont{H.}~\bibnamefont{Kawano}},
  \bibinfo{author}{\bibfnamefont{R.}~\bibnamefont{Kajimoto}},
  \bibinfo{author}{\bibfnamefont{H.}~\bibnamefont{Yoshizawa}},
  \bibinfo{author}{\bibfnamefont{J.}~\bibnamefont{Fernandez-Baca}},
  \bibinfo{author}{\bibfnamefont{Y.}~\bibnamefont{Tomioka}},
  \bibinfo{author}{\bibfnamefont{H.}~\bibnamefont{Kuwahara}}, \bibnamefont{and}
  \bibinfo{author}{\bibfnamefont{Y.}~\bibnamefont{Tokura}},
  \bibinfo{journal}{Physica} \textbf{\bibinfo{volume}{289}}
  (\bibinfo{year}{1998}).

\bibitem[{\citenamefont{van~den Brink et~al.}(1999)\citenamefont{van~den Brink,
  Khaliullin, and Khomskii}}]{Brink+99}
\bibinfo{author}{\bibfnamefont{J.}~\bibnamefont{van~den Brink}},
  \bibinfo{author}{\bibfnamefont{G.}~\bibnamefont{Khaliullin}},
  \bibnamefont{and} \bibinfo{author}{\bibfnamefont{D.}~\bibnamefont{Khomskii}},
  \bibinfo{journal}{Phys. Rev. Lett.} \textbf{\bibinfo{volume}{83}},
  \bibinfo{pages}{5118} (\bibinfo{year}{1999}).

\bibitem[{\citenamefont{Yunoki et~al.}(2000)\citenamefont{Yunoki, Hotta, and
  Dagotto}}]{Yunoki+2000}
\bibinfo{author}{\bibfnamefont{S.}~\bibnamefont{Yunoki}},
  \bibinfo{author}{\bibfnamefont{T.}~\bibnamefont{Hotta}}, \bibnamefont{and}
  \bibinfo{author}{\bibfnamefont{E.}~\bibnamefont{Dagotto}},
  \bibinfo{journal}{Phys. Rev. Lett.} \textbf{\bibinfo{volume}{84}},
  \bibinfo{pages}{3714} (\bibinfo{year}{2000}).

\bibitem[{\citenamefont{Kajimoto et~al.}(2002)\citenamefont{Kajimoto,
  Yoshizawa, Tomioka, and Tokura}}]{Kajimoto+02}
\bibinfo{author}{\bibfnamefont{R.}~\bibnamefont{Kajimoto}},
  \bibinfo{author}{\bibfnamefont{H.}~\bibnamefont{Yoshizawa}},
  \bibinfo{author}{\bibfnamefont{Y.}~\bibnamefont{Tomioka}}, \bibnamefont{and}
  \bibinfo{author}{\bibfnamefont{Y.}~\bibnamefont{Tokura}},
  \bibinfo{journal}{Phys. Rev. B} \textbf{\bibinfo{volume}{66}},
  \bibinfo{pages}{180402} (\bibinfo{year}{2002}).

\bibitem[{\citenamefont{Wollan and Koehler}(1955)}]{Wollan+55}
\bibinfo{author}{\bibfnamefont{E.~O.} \bibnamefont{Wollan}} \bibnamefont{and}
  \bibinfo{author}{\bibfnamefont{W.~C.} \bibnamefont{Koehler}},
  \bibinfo{journal}{Phys. Rev.} \textbf{\bibinfo{volume}{100}},
  \bibinfo{pages}{545} (\bibinfo{year}{1955}).

\bibitem[{\citenamefont{Goodenough}(1955)}]{Gooden+55}
\bibinfo{author}{\bibfnamefont{J.}~\bibnamefont{Goodenough}},
  \bibinfo{journal}{Phys. Rev.} \textbf{\bibinfo{volume}{200}},
  \bibinfo{pages}{564} (\bibinfo{year}{1955}).

\bibitem[{\citenamefont{Jir\'ak et~al.}(2000)\citenamefont{Jir\'ak, Damay,
  Hervieu, Martin, Raveau, Andr\'e, and Bour\'ee}}]{Jirak+00}
\bibinfo{author}{\bibfnamefont{Z.}~\bibnamefont{Jir\'ak}},
  \bibinfo{author}{\bibfnamefont{F.}~\bibnamefont{Damay}},
  \bibinfo{author}{\bibfnamefont{M.}~\bibnamefont{Hervieu}},
  \bibinfo{author}{\bibfnamefont{C.}~\bibnamefont{Martin}},
  \bibinfo{author}{\bibfnamefont{B.}~\bibnamefont{Raveau}},
  \bibinfo{author}{\bibfnamefont{G.}~\bibnamefont{Andr\'e}}, \bibnamefont{and}
  \bibinfo{author}{\bibfnamefont{F.}~\bibnamefont{Bour\'ee}},
  \bibinfo{journal}{Phys. Rev. B} \textbf{\bibinfo{volume}{61}},
  \bibinfo{pages}{1181} (\bibinfo{year}{2000}).

\bibitem[{\citenamefont{Herrero-Mart\'\i{}n
  et~al.}(2004)\citenamefont{Herrero-Mart\'\i{}n, Garc\'\i{}a, Sub\'\i{}as,
  Blasco, and S\'anchez}}]{Herrero+04}
\bibinfo{author}{\bibfnamefont{J.}~\bibnamefont{Herrero-Mart\'\i{}n}},
  \bibinfo{author}{\bibfnamefont{J.}~\bibnamefont{Garc\'\i{}a}},
  \bibinfo{author}{\bibfnamefont{G.}~\bibnamefont{Sub\'\i{}as}},
  \bibinfo{author}{\bibfnamefont{J.}~\bibnamefont{Blasco}}, \bibnamefont{and}
  \bibinfo{author}{\bibfnamefont{M.~C.} \bibnamefont{S\'anchez}},
  \bibinfo{journal}{Phys. Rev. B} \textbf{\bibinfo{volume}{70}},
  \bibinfo{pages}{024408} (\bibinfo{year}{2004}).

\bibitem[{\citenamefont{Doud-Aladine et~al.}(2002)\citenamefont{Doud-Aladine,
  Rodriguez-Carvajal, L.Pinsard-Gaudart, Fernandez-Diaz, and
  Revcolevschi}}]{doud+02}
\bibinfo{author}{\bibfnamefont{A.}~\bibnamefont{Doud-Aladine}},
  \bibinfo{author}{\bibfnamefont{J.}~\bibnamefont{Rodriguez-Carvajal}},
  \bibinfo{author}{\bibnamefont{L.Pinsard-Gaudart}},
  \bibinfo{author}{\bibfnamefont{M.}~\bibnamefont{Fernandez-Diaz}},
  \bibnamefont{and}
  \bibinfo{author}{\bibfnamefont{A.}~\bibnamefont{Revcolevschi}},
  \bibinfo{journal}{Phys. Rev. Lett.} \textbf{\bibinfo{volume}{89}},
  \bibinfo{pages}{097205} (\bibinfo{year}{2002}).

\bibitem[{\citenamefont{Efremov et~al.}(2004)\citenamefont{Efremov, van~den
  Brink, and Khomskii}}]{Efremov+04}
\bibinfo{author}{\bibfnamefont{D.}~\bibnamefont{Efremov}},
  \bibinfo{author}{\bibfnamefont{J.}~\bibnamefont{van~den Brink}},
  \bibnamefont{and} \bibinfo{author}{\bibfnamefont{D.}~\bibnamefont{Khomskii}},
  \bibinfo{journal}{Nature Mater.} \textbf{\bibinfo{volume}{3}},
  \bibinfo{pages}{853} (\bibinfo{year}{2004}).

\bibitem[{\citenamefont{Senff et~al.}(2006)\citenamefont{Senff, Kr\"{u}ger,
  Scheidl, Benomar, Sidis, Demmel, and Braden}}]{Senff+06}
\bibinfo{author}{\bibfnamefont{D.}~\bibnamefont{Senff}},
  \bibinfo{author}{\bibfnamefont{F.}~\bibnamefont{Kr\"{u}ger}},
  \bibinfo{author}{\bibfnamefont{S.}~\bibnamefont{Scheidl}},
  \bibinfo{author}{\bibfnamefont{M.}~\bibnamefont{Benomar}},
  \bibinfo{author}{\bibfnamefont{Y.}~\bibnamefont{Sidis}},
  \bibinfo{author}{\bibfnamefont{F.}~\bibnamefont{Demmel}}, \bibnamefont{and}
  \bibinfo{author}{\bibfnamefont{M.}~\bibnamefont{Braden}},
  \bibinfo{journal}{Phys. Rev. Lett.} \textbf{\bibinfo{volume}{96}},
  \bibinfo{pages}{257201} (\bibinfo{year}{2006}).

\bibitem[{\citenamefont{Jung et~al.}(2005)\citenamefont{Jung, Markovich,
  van~der Beek, Mogilyansky, and Mukovskii}}]{Jung+05}
\bibinfo{author}{\bibfnamefont{G.}~\bibnamefont{Jung}},
  \bibinfo{author}{\bibfnamefont{V.}~\bibnamefont{Markovich}},
  \bibinfo{author}{\bibfnamefont{C.~J.} \bibnamefont{van~der Beek}},
  \bibinfo{author}{\bibfnamefont{D.}~\bibnamefont{Mogilyansky}},
  \bibnamefont{and} \bibinfo{author}{\bibfnamefont{Y.~M.}
  \bibnamefont{Mukovskii}}, \bibinfo{journal}{Phys. Rev. B}
  \textbf{\bibinfo{volume}{72}}, \bibinfo{pages}{134412}
  (\bibinfo{year}{2005}).

\bibitem[{\citenamefont{Reutler et~al.}(2003)\citenamefont{Reutler, Friedt,
  Büchner, Braden, and Revcolevschi}}]{Reutler+03}
\bibinfo{author}{\bibfnamefont{P.}~\bibnamefont{Reutler}},
  \bibinfo{author}{\bibfnamefont{O.}~\bibnamefont{Friedt}},
  \bibinfo{author}{\bibfnamefont{B.}~\bibnamefont{Büchner}},
  \bibinfo{author}{\bibfnamefont{M.}~\bibnamefont{Braden}}, \bibnamefont{and}
  \bibinfo{author}{\bibfnamefont{A.}~\bibnamefont{Revcolevschi}},
  \bibinfo{journal}{Journal of Crystal Growth} \textbf{\bibinfo{volume}{249}},
  \bibinfo{pages}{222 } (\bibinfo{year}{2003}).

\bibitem[{\citenamefont{Senff et~al.}(2005)\citenamefont{Senff, Reutler,
  Braden, Friedt, Bruns, Cousson, Bour\'ee, Merz, B\"uchner, and
  Revcolevschi}}]{Senff+05}
\bibinfo{author}{\bibfnamefont{D.}~\bibnamefont{Senff}},
  \bibinfo{author}{\bibfnamefont{P.}~\bibnamefont{Reutler}},
  \bibinfo{author}{\bibfnamefont{M.}~\bibnamefont{Braden}},
  \bibinfo{author}{\bibfnamefont{O.}~\bibnamefont{Friedt}},
  \bibinfo{author}{\bibfnamefont{D.}~\bibnamefont{Bruns}},
  \bibinfo{author}{\bibfnamefont{A.}~\bibnamefont{Cousson}},
  \bibinfo{author}{\bibfnamefont{F.}~\bibnamefont{Bour\'ee}},
  \bibinfo{author}{\bibfnamefont{M.}~\bibnamefont{Merz}},
  \bibinfo{author}{\bibfnamefont{B.}~\bibnamefont{B\"uchner}},
  \bibnamefont{and}
  \bibinfo{author}{\bibfnamefont{A.}~\bibnamefont{Revcolevschi}},
  \bibinfo{journal}{Phys. Rev. B} \textbf{\bibinfo{volume}{71}},
  \bibinfo{pages}{024425} (\bibinfo{year}{2005}).

\bibitem[{\citenamefont{Senff et~al.}(2008)\citenamefont{Senff, Schumann,
  Benomar, Kriener, Lorenz, Sidis, Habicht, Link, and Braden}}]{Senff+08}
\bibinfo{author}{\bibfnamefont{D.}~\bibnamefont{Senff}},
  \bibinfo{author}{\bibfnamefont{O.}~\bibnamefont{Schumann}},
  \bibinfo{author}{\bibfnamefont{M.}~\bibnamefont{Benomar}},
  \bibinfo{author}{\bibfnamefont{M.}~\bibnamefont{Kriener}},
  \bibinfo{author}{\bibfnamefont{T.}~\bibnamefont{Lorenz}},
  \bibinfo{author}{\bibfnamefont{Y.}~\bibnamefont{Sidis}},
  \bibinfo{author}{\bibfnamefont{K.}~\bibnamefont{Habicht}},
  \bibinfo{author}{\bibfnamefont{P.}~\bibnamefont{Link}}, \bibnamefont{and}
  \bibinfo{author}{\bibfnamefont{M.}~\bibnamefont{Braden}},
  \bibinfo{journal}{Phys. Rev. B} \textbf{\bibinfo{volume}{77}},
  \bibinfo{pages}{184413} (\bibinfo{year}{2008}).

\bibitem[{\citenamefont{Ulbrich et~al.}(2011)\citenamefont{Ulbrich, Senff,
  Steffens, Schumann, Sidis, Reutler, Revcolevschi, and Braden}}]{Ulbrich+11}
\bibinfo{author}{\bibfnamefont{H.}~\bibnamefont{Ulbrich}},
  \bibinfo{author}{\bibfnamefont{D.}~\bibnamefont{Senff}},
  \bibinfo{author}{\bibfnamefont{P.}~\bibnamefont{Steffens}},
  \bibinfo{author}{\bibfnamefont{O.~J.} \bibnamefont{Schumann}},
  \bibinfo{author}{\bibfnamefont{Y.}~\bibnamefont{Sidis}},
  \bibinfo{author}{\bibfnamefont{P.}~\bibnamefont{Reutler}},
  \bibinfo{author}{\bibfnamefont{A.}~\bibnamefont{Revcolevschi}},
  \bibnamefont{and} \bibinfo{author}{\bibfnamefont{M.}~\bibnamefont{Braden}},
  \bibinfo{journal}{Phys. Rev. Lett.} \textbf{\bibinfo{volume}{106}},
  \bibinfo{pages}{157201} (\bibinfo{year}{2011}).

\bibitem[{\citenamefont{Tokura and Nagaosa}(2000)}]{Tokura+00}
\bibinfo{author}{\bibfnamefont{Y.}~\bibnamefont{Tokura}} \bibnamefont{and}
  \bibinfo{author}{\bibfnamefont{N.}~\bibnamefont{Nagaosa}},
  \bibinfo{journal}{Science} \textbf{\bibinfo{volume}{288}},
  \bibinfo{pages}{462} (\bibinfo{year}{2000}).

\bibitem[{\citenamefont{Tokunaga et~al.}(1999)\citenamefont{Tokunaga, Miura,
  Moritoma, and Tokura}}]{Tokunaga+99}
\bibinfo{author}{\bibfnamefont{M.}~\bibnamefont{Tokunaga}},
  \bibinfo{author}{\bibfnamefont{N.}~\bibnamefont{Miura}},
  \bibinfo{author}{\bibfnamefont{Y.}~\bibnamefont{Moritoma}}, \bibnamefont{and}
  \bibinfo{author}{\bibfnamefont{Y.}~\bibnamefont{Tokura}},
  \bibinfo{journal}{Phys. Rev. B} \textbf{\bibinfo{volume}{59}},
  \bibinfo{pages}{11151} (\bibinfo{year}{1999}).

\bibitem[{\citenamefont{Yoshizawa et~al.}(1998)\citenamefont{Yoshizawa, Kawano,
  Fernandez-Baca, Kuwahara, and Tokura}}]{Yoshizawa+98}
\bibinfo{author}{\bibfnamefont{H.}~\bibnamefont{Yoshizawa}},
  \bibinfo{author}{\bibfnamefont{H.}~\bibnamefont{Kawano}},
  \bibinfo{author}{\bibfnamefont{J.}~\bibnamefont{Fernandez-Baca}},
  \bibinfo{author}{\bibfnamefont{H.}~\bibnamefont{Kuwahara}}, \bibnamefont{and}
  \bibinfo{author}{\bibfnamefont{Y.}~\bibnamefont{Tokura}},
  \bibinfo{journal}{Phys. Rev. B} \textbf{\bibinfo{volume}{58}}
  (\bibinfo{year}{1998}).

\bibitem[{\citenamefont{Fernandez-Baca
  et~al.}(1998)\citenamefont{Fernandez-Baca, Dai, Hwang, Kloc, and
  Cheong}}]{Fernandez+98}
\bibinfo{author}{\bibfnamefont{J.~A.} \bibnamefont{Fernandez-Baca}},
  \bibinfo{author}{\bibfnamefont{P.}~\bibnamefont{Dai}},
  \bibinfo{author}{\bibfnamefont{H.~Y.} \bibnamefont{Hwang}},
  \bibinfo{author}{\bibfnamefont{C.}~\bibnamefont{Kloc}}, \bibnamefont{and}
  \bibinfo{author}{\bibfnamefont{S.-W.} \bibnamefont{Cheong}},
  \bibinfo{journal}{Phys. Rev. Lett.} \textbf{\bibinfo{volume}{80}},
  \bibinfo{pages}{4012} (\bibinfo{year}{1998}).

\bibitem[{\citenamefont{Kawano-Furukawa
  et~al.}(2003)\citenamefont{Kawano-Furukawa, Kajimoto, Yoshizawa, Tomioka,
  Kuwahara, and Tokura}}]{Kawano-F+03}
\bibinfo{author}{\bibfnamefont{H.}~\bibnamefont{Kawano-Furukawa}},
  \bibinfo{author}{\bibfnamefont{R.}~\bibnamefont{Kajimoto}},
  \bibinfo{author}{\bibfnamefont{H.}~\bibnamefont{Yoshizawa}},
  \bibinfo{author}{\bibfnamefont{Y.}~\bibnamefont{Tomioka}},
  \bibinfo{author}{\bibfnamefont{H.}~\bibnamefont{Kuwahara}}, \bibnamefont{and}
  \bibinfo{author}{\bibfnamefont{Y.}~\bibnamefont{Tokura}},
  \bibinfo{journal}{Phys. Rev. B} \textbf{\bibinfo{volume}{67}},
  \bibinfo{pages}{174422} (\bibinfo{year}{2003}).

\bibitem[{\citenamefont{Endoh et~al.}(2005)\citenamefont{Endoh, Hiraka,
  Tomioka, Tokura, Nagaosa, and Fujiwara}}]{Endoh+05}
\bibinfo{author}{\bibfnamefont{Y.}~\bibnamefont{Endoh}},
  \bibinfo{author}{\bibfnamefont{H.}~\bibnamefont{Hiraka}},
  \bibinfo{author}{\bibfnamefont{Y.}~\bibnamefont{Tomioka}},
  \bibinfo{author}{\bibfnamefont{Y.}~\bibnamefont{Tokura}},
  \bibinfo{author}{\bibfnamefont{N.}~\bibnamefont{Nagaosa}}, \bibnamefont{and}
  \bibinfo{author}{\bibfnamefont{T.}~\bibnamefont{Fujiwara}},
  \bibinfo{journal}{Phys. Rev. Lett.} \textbf{\bibinfo{volume}{94}},
  \bibinfo{pages}{017206} (\bibinfo{year}{2005}).

\bibitem[{\citenamefont{Hennion et~al.}(2005)\citenamefont{Hennion, Moussa,
  Lehouelleur, Wang, Ivanov, Mukovskii, and Shulyatev}}]{Hennion+05}
\bibinfo{author}{\bibfnamefont{M.}~\bibnamefont{Hennion}},
  \bibinfo{author}{\bibfnamefont{F.}~\bibnamefont{Moussa}},
  \bibinfo{author}{\bibfnamefont{P.}~\bibnamefont{Lehouelleur}},
  \bibinfo{author}{\bibfnamefont{F.}~\bibnamefont{Wang}},
  \bibinfo{author}{\bibfnamefont{A.}~\bibnamefont{Ivanov}},
  \bibinfo{author}{\bibfnamefont{Y.~M.} \bibnamefont{Mukovskii}},
  \bibnamefont{and}
  \bibinfo{author}{\bibfnamefont{D.}~\bibnamefont{Shulyatev}},
  \bibinfo{journal}{Phys. Rev. Lett.} \textbf{\bibinfo{volume}{94}},
  \bibinfo{pages}{057006} (\bibinfo{year}{2005}).

\bibitem[{\citenamefont{Kirby et~al.}(2006)\citenamefont{Kirby, Rhyne, Kaiser,
  Kuwahara, and Tokura}}]{Kirby+06}
\bibinfo{author}{\bibfnamefont{B.}~\bibnamefont{Kirby}},
  \bibinfo{author}{\bibfnamefont{J.}~\bibnamefont{Rhyne}},
  \bibinfo{author}{\bibfnamefont{H.}~\bibnamefont{Kaiser}},
  \bibinfo{author}{\bibfnamefont{H.}~\bibnamefont{Kuwahara}}, \bibnamefont{and}
  \bibinfo{author}{\bibfnamefont{Y.}~\bibnamefont{Tokura}},
  \bibinfo{journal}{J. Magn. Magn. Mater.} \textbf{\bibinfo{volume}{302}},
  \bibinfo{pages}{237} (\bibinfo{year}{2006}).

\bibitem[{\citenamefont{Ye et~al.}(2006)\citenamefont{Ye, Dai, Fernandez-Baca,
  Sha, Lynn, Kawano-Furukawa, Tomioka, Tokura, and Zhang}}]{Ye+06}
\bibinfo{author}{\bibfnamefont{F.}~\bibnamefont{Ye}},
  \bibinfo{author}{\bibfnamefont{P.}~\bibnamefont{Dai}},
  \bibinfo{author}{\bibfnamefont{J.~A.} \bibnamefont{Fernandez-Baca}},
  \bibinfo{author}{\bibfnamefont{H.}~\bibnamefont{Sha}},
  \bibinfo{author}{\bibfnamefont{J.~W.} \bibnamefont{Lynn}},
  \bibinfo{author}{\bibfnamefont{H.}~\bibnamefont{Kawano-Furukawa}},
  \bibinfo{author}{\bibfnamefont{Y.}~\bibnamefont{Tomioka}},
  \bibinfo{author}{\bibfnamefont{Y.}~\bibnamefont{Tokura}}, \bibnamefont{and}
  \bibinfo{author}{\bibfnamefont{J.}~\bibnamefont{Zhang}},
  \bibinfo{journal}{Phys. Rev. Lett.} \textbf{\bibinfo{volume}{96}},
  \bibinfo{pages}{047204} (\bibinfo{year}{2006}).

\bibitem[{\citenamefont{Ye et~al.}(2007)\citenamefont{Ye, Dai, Fernandez-Baca,
  Adroja, Perring, Tomioka, and Tokura}}]{Ye+07}
\bibinfo{author}{\bibfnamefont{F.}~\bibnamefont{Ye}},
  \bibinfo{author}{\bibfnamefont{P.}~\bibnamefont{Dai}},
  \bibinfo{author}{\bibfnamefont{J.~A.} \bibnamefont{Fernandez-Baca}},
  \bibinfo{author}{\bibfnamefont{D.~T.} \bibnamefont{Adroja}},
  \bibinfo{author}{\bibfnamefont{T.~G.} \bibnamefont{Perring}},
  \bibinfo{author}{\bibfnamefont{Y.}~\bibnamefont{Tomioka}}, \bibnamefont{and}
  \bibinfo{author}{\bibfnamefont{Y.}~\bibnamefont{Tokura}},
  \bibinfo{journal}{Phys. Rev. B} \textbf{\bibinfo{volume}{75}},
  \bibinfo{pages}{144408} (\bibinfo{year}{2007}).

\bibitem[{\citenamefont{Petit et~al.}(2009)\citenamefont{Petit, Hennion,
  Moussa, Lamago, Ivanov, Mukovskii, and Shulyatev}}]{Petit+09}
\bibinfo{author}{\bibfnamefont{S.}~\bibnamefont{Petit}},
  \bibinfo{author}{\bibfnamefont{M.}~\bibnamefont{Hennion}},
  \bibinfo{author}{\bibfnamefont{F.}~\bibnamefont{Moussa}},
  \bibinfo{author}{\bibfnamefont{D.}~\bibnamefont{Lamago}},
  \bibinfo{author}{\bibfnamefont{A.}~\bibnamefont{Ivanov}},
  \bibinfo{author}{\bibfnamefont{Y.~M.} \bibnamefont{Mukovskii}},
  \bibnamefont{and}
  \bibinfo{author}{\bibfnamefont{D.}~\bibnamefont{Shulyatev}},
  \bibinfo{journal}{Phys. Rev. Lett.} \textbf{\bibinfo{volume}{102}},
  \bibinfo{pages}{207201} (\bibinfo{year}{2009}).

\bibitem[{\citenamefont{Kajimoto et~al.}(1999)\citenamefont{Kajimoto,
  Yoshizawa, Kawano, Kuwahara, Tokura, Ohoyama, and Ohashi}}]{Kajimoto+99}
\bibinfo{author}{\bibfnamefont{R.}~\bibnamefont{Kajimoto}},
  \bibinfo{author}{\bibfnamefont{H.}~\bibnamefont{Yoshizawa}},
  \bibinfo{author}{\bibfnamefont{H.}~\bibnamefont{Kawano}},
  \bibinfo{author}{\bibfnamefont{H.}~\bibnamefont{Kuwahara}},
  \bibinfo{author}{\bibfnamefont{Y.}~\bibnamefont{Tokura}},
  \bibinfo{author}{\bibfnamefont{K.}~\bibnamefont{Ohoyama}}, \bibnamefont{and}
  \bibinfo{author}{\bibfnamefont{M.}~\bibnamefont{Ohashi}},
  \bibinfo{journal}{Phys. Rev. B} \textbf{\bibinfo{volume}{60}},
  \bibinfo{pages}{9506} (\bibinfo{year}{1999}).

\bibitem[{\citenamefont{Tokura}(2006)}]{Tokura+06}
\bibinfo{author}{\bibfnamefont{Y.}~\bibnamefont{Tokura}},
  \bibinfo{journal}{Rep. Prog. Phys.} \textbf{\bibinfo{volume}{69}},
  \bibinfo{pages}{797} (\bibinfo{year}{2006}).

\bibitem[{\citenamefont{Kuwahara et~al.}(1995)\citenamefont{Kuwahara, Tomioka,
  Asamitsu, Moritomo, and Tokura}}]{Kuwahara+95}
\bibinfo{author}{\bibfnamefont{H.}~\bibnamefont{Kuwahara}},
  \bibinfo{author}{\bibfnamefont{Y.}~\bibnamefont{Tomioka}},
  \bibinfo{author}{\bibfnamefont{A.}~\bibnamefont{Asamitsu}},
  \bibinfo{author}{\bibfnamefont{Y.}~\bibnamefont{Moritomo}}, \bibnamefont{and}
  \bibinfo{author}{\bibfnamefont{Y.}~\bibnamefont{Tokura}},
  \bibinfo{journal}{Science} \textbf{\bibinfo{volume}{270}},
  \bibinfo{pages}{961} (\bibinfo{year}{1995}).

\bibitem[{\citenamefont{Kawano et~al.}(1997)\citenamefont{Kawano, Kajimoto,
  Yoshizawa, Tomioka, Kuwahara, and Tokura}}]{Kawano+97}
\bibinfo{author}{\bibfnamefont{H.}~\bibnamefont{Kawano}},
  \bibinfo{author}{\bibfnamefont{R.}~\bibnamefont{Kajimoto}},
  \bibinfo{author}{\bibfnamefont{H.}~\bibnamefont{Yoshizawa}},
  \bibinfo{author}{\bibfnamefont{Y.}~\bibnamefont{Tomioka}},
  \bibinfo{author}{\bibfnamefont{H.}~\bibnamefont{Kuwahara}}, \bibnamefont{and}
  \bibinfo{author}{\bibfnamefont{Y.}~\bibnamefont{Tokura}},
  \bibinfo{journal}{Phys. Rev. Lett.} \textbf{\bibinfo{volume}{78}},
  \bibinfo{pages}{4253} (\bibinfo{year}{1997}).

\bibitem[{\citenamefont{Rodr\'\i{}guez-Carvajal}(1993)}]{Carvajal+93}
\bibinfo{author}{\bibfnamefont{J.}~\bibnamefont{Rodr\'\i{}guez-Carvajal}},
  \bibinfo{journal}{Physica B} \textbf{\bibinfo{volume}{192}},
  \bibinfo{pages}{55} (\bibinfo{year}{1993}).

\bibitem[{Imm()}]{Imma}
\emph{\bibinfo{title}{In space group {I}mma it would align along the
  orthorhombic b-direction.}}

\bibitem[{\citenamefont{Petitgrand et~al.}(1997)\citenamefont{Petitgrand,
  Casalta, Bourges, and Ivanov}}]{Petitgrand+97}
\bibinfo{author}{\bibfnamefont{D.}~\bibnamefont{Petitgrand}},
  \bibinfo{author}{\bibfnamefont{H.}~\bibnamefont{Casalta}},
  \bibinfo{author}{\bibfnamefont{P.}~\bibnamefont{Bourges}}, \bibnamefont{and}
  \bibinfo{author}{\bibfnamefont{A.}~\bibnamefont{Ivanov}},
  \bibinfo{journal}{J., Physica B} \textbf{\bibinfo{volume}{806}},
  \bibinfo{pages}{234} (\bibinfo{year}{1997}).

\bibitem[{\citenamefont{Krishnamurthy et~al.}(2006)\citenamefont{Krishnamurthy,
  Robertson, Fishman, Lumsden, and Mitchell}}]{Krishnamurthy+06}
\bibinfo{author}{\bibfnamefont{V.~V.} \bibnamefont{Krishnamurthy}},
  \bibinfo{author}{\bibfnamefont{J.~L.} \bibnamefont{Robertson}},
  \bibinfo{author}{\bibfnamefont{R.~S.} \bibnamefont{Fishman}},
  \bibinfo{author}{\bibfnamefont{M.~D.} \bibnamefont{Lumsden}},
  \bibnamefont{and} \bibinfo{author}{\bibfnamefont{J.~F.}
  \bibnamefont{Mitchell}}, \bibinfo{journal}{Phys. Rev. B}
  \textbf{\bibinfo{volume}{73}}, \bibinfo{pages}{060404}
  (\bibinfo{year}{2006}).

\bibitem[{\citenamefont{Feiner and Oles}(1999)}]{Feiner+99}
\bibinfo{author}{\bibfnamefont{L.~F.} \bibnamefont{Feiner}} \bibnamefont{and}
  \bibinfo{author}{\bibfnamefont{A.~M.} \bibnamefont{Oles}},
  \bibinfo{journal}{Phys. Rev. B} \textbf{\bibinfo{volume}{59}},
  \bibinfo{pages}{3295} (\bibinfo{year}{1999}).

\bibitem[{\citenamefont{Hotta et~al.}(1999)\citenamefont{Hotta, Yunoki, Mayr,
  and Dagotto}}]{Hotta+99}
\bibinfo{author}{\bibfnamefont{T.}~\bibnamefont{Hotta}},
  \bibinfo{author}{\bibfnamefont{S.}~\bibnamefont{Yunoki}},
  \bibinfo{author}{\bibfnamefont{M.}~\bibnamefont{Mayr}}, \bibnamefont{and}
  \bibinfo{author}{\bibfnamefont{E.}~\bibnamefont{Dagotto}},
  \bibinfo{journal}{Phys. Rev. B} \textbf{\bibinfo{volume}{60}},
  \bibinfo{pages}{R15009} (\bibinfo{year}{1999}).

\bibitem[{\citenamefont{Sternlieb et~al.}(1996)\citenamefont{Sternlieb, Hill,
  Wildgruber, Luke, Nachumi, Moritomo, and Tokura}}]{Sternlieb+96}
\bibinfo{author}{\bibfnamefont{B.~J.} \bibnamefont{Sternlieb}},
  \bibinfo{author}{\bibfnamefont{J.~P.} \bibnamefont{Hill}},
  \bibinfo{author}{\bibfnamefont{U.~C.} \bibnamefont{Wildgruber}},
  \bibinfo{author}{\bibfnamefont{G.~M.} \bibnamefont{Luke}},
  \bibinfo{author}{\bibfnamefont{B.}~\bibnamefont{Nachumi}},
  \bibinfo{author}{\bibfnamefont{Y.}~\bibnamefont{Moritomo}}, \bibnamefont{and}
  \bibinfo{author}{\bibfnamefont{Y.}~\bibnamefont{Tokura}},
  \bibinfo{journal}{Phys. Rev. Lett.} \textbf{\bibinfo{volume}{76}},
  \bibinfo{pages}{2169} (\bibinfo{year}{1996}).

\bibitem[{\citenamefont{Shannon and Chubukov}(2002)}]{Shannon+02}
\bibinfo{author}{\bibfnamefont{N.}~\bibnamefont{Shannon}} \bibnamefont{and}
  \bibinfo{author}{\bibfnamefont{A.~V.} \bibnamefont{Chubukov}},
  \bibinfo{journal}{Phys. Rev. B} \textbf{\bibinfo{volume}{65}},
  \bibinfo{pages}{104418} (\bibinfo{year}{2002}).

\bibitem[{\citenamefont{Lv et~al.}(2010)\citenamefont{Lv, Kr\"uger, and
  Phillips}}]{Lv+10}
\bibinfo{author}{\bibfnamefont{W.}~\bibnamefont{Lv}},
  \bibinfo{author}{\bibfnamefont{F.}~\bibnamefont{Kr\"uger}}, \bibnamefont{and}
  \bibinfo{author}{\bibfnamefont{P.}~\bibnamefont{Phillips}},
  \bibinfo{journal}{Phys. Rev. B} \textbf{\bibinfo{volume}{82}},
  \bibinfo{pages}{045125} (\bibinfo{year}{2010}).

\bibitem[{\citenamefont{Ventura and Alascio}(2003)}]{Ventura+03}
\bibinfo{author}{\bibfnamefont{C.~I.} \bibnamefont{Ventura}} \bibnamefont{and}
  \bibinfo{author}{\bibfnamefont{B.}~\bibnamefont{Alascio}},
  \bibinfo{journal}{Phys. Rev. B} \textbf{\bibinfo{volume}{68}},
  \bibinfo{pages}{R020404} (\bibinfo{year}{2003}).

\bibitem[{\citenamefont{Sikora and Ole\'{s}}(2005)}]{Sikora+05}
\bibinfo{author}{\bibfnamefont{O.}~\bibnamefont{Sikora}} \bibnamefont{and}
  \bibinfo{author}{\bibfnamefont{A.~M.} \bibnamefont{Ole\'{s}}},
  \bibinfo{journal}{Physica B} \textbf{\bibinfo{volume}{359-361}},
  \bibinfo{pages}{1300} (\bibinfo{year}{2005}).

\bibitem[{\citenamefont{Gontchar and Nikiforov}(2002)}]{Gontchar+02}
\bibinfo{author}{\bibfnamefont{L.~E.} \bibnamefont{Gontchar}} \bibnamefont{and}
  \bibinfo{author}{\bibfnamefont{A.~E.} \bibnamefont{Nikiforov}},
  \bibinfo{journal}{Phys. Rev. B} \textbf{\bibinfo{volume}{66}},
  \bibinfo{pages}{014437} (\bibinfo{year}{2002}).

\bibitem[{\citenamefont{Jongh}(1990)}]{Jongh+90}
\bibinfo{author}{\bibfnamefont{L.}~\bibnamefont{Jongh}},
  \emph{\bibinfo{title}{Magnetic properties of layered transition metal
  compounds}} (\bibinfo{publisher}{Kluwer Academic}, \bibinfo{year}{1990}).

\bibitem[{\citenamefont{Perring et~al.}(1996)\citenamefont{Perring, Aeppli,
  Hayden, Carter, Remeika, and Cheong}}]{Perring+96}
\bibinfo{author}{\bibfnamefont{T.~G.} \bibnamefont{Perring}},
  \bibinfo{author}{\bibfnamefont{G.}~\bibnamefont{Aeppli}},
  \bibinfo{author}{\bibfnamefont{S.~M.} \bibnamefont{Hayden}},
  \bibinfo{author}{\bibfnamefont{S.~A.} \bibnamefont{Carter}},
  \bibinfo{author}{\bibfnamefont{J.~P.} \bibnamefont{Remeika}},
  \bibnamefont{and} \bibinfo{author}{\bibfnamefont{S.-W.}
  \bibnamefont{Cheong}}, \bibinfo{journal}{Phys. Rev. Lett.}
  \textbf{\bibinfo{volume}{77}}, \bibinfo{pages}{711} (\bibinfo{year}{1996}).

\bibitem[{\citenamefont{Hwang et~al.}(1998)\citenamefont{Hwang, Dai, Cheong,
  Aeppli, Tennant, and Mook}}]{Hwang+98}
\bibinfo{author}{\bibfnamefont{H.}~\bibnamefont{Hwang}},
  \bibinfo{author}{\bibfnamefont{P.}~\bibnamefont{Dai}},
  \bibinfo{author}{\bibfnamefont{S.}~\bibnamefont{Cheong}},
  \bibinfo{author}{\bibfnamefont{G.}~\bibnamefont{Aeppli}},
  \bibinfo{author}{\bibfnamefont{D.}~\bibnamefont{Tennant}}, \bibnamefont{and}
  \bibinfo{author}{\bibfnamefont{H.}~\bibnamefont{Mook}},
  \bibinfo{journal}{Phys. Rev. Lett.} \textbf{\bibinfo{volume}{80}},
  \bibinfo{pages}{1316} (\bibinfo{year}{1998}).

\bibitem[{\citenamefont{Zhang et~al.}(2007)\citenamefont{Zhang, Ye, Sha, Dai,
  Fernandez-Baca, and Plummer}}]{Zhang+07}
\bibinfo{author}{\bibfnamefont{J.}~\bibnamefont{Zhang}},
  \bibinfo{author}{\bibfnamefont{F.}~\bibnamefont{Ye}},
  \bibinfo{author}{\bibfnamefont{H.}~\bibnamefont{Sha}},
  \bibinfo{author}{\bibfnamefont{P.}~\bibnamefont{Dai}},
  \bibinfo{author}{\bibfnamefont{J.~A.} \bibnamefont{Fernandez-Baca}},
  \bibnamefont{and} \bibinfo{author}{\bibfnamefont{E.~W.}
  \bibnamefont{Plummer}}, \bibinfo{journal}{Journal of Physics: Condensed
  Matter} \textbf{\bibinfo{volume}{19}}, \bibinfo{pages}{315204}
  (\bibinfo{year}{2007}).

\bibitem[{\citenamefont{Zheludev}(2006)}]{Reslib}
\bibinfo{author}{\bibfnamefont{A.}~\bibnamefont{Zheludev}},
  \emph{\bibinfo{title}{ResLib 3.3}} (\bibinfo{publisher}{Oak Ridge National
  Laboratory, Oak Ridge, 2006}, \bibinfo{year}{2006}).

\bibitem[{\citenamefont{Martin et~al.}(1996)\citenamefont{Martin, Shirane,
  Endoh, Hirota, Moritomo, and Tokura}}]{Martin+96}
\bibinfo{author}{\bibfnamefont{M.~C.} \bibnamefont{Martin}},
  \bibinfo{author}{\bibfnamefont{G.}~\bibnamefont{Shirane}},
  \bibinfo{author}{\bibfnamefont{Y.}~\bibnamefont{Endoh}},
  \bibinfo{author}{\bibfnamefont{K.}~\bibnamefont{Hirota}},
  \bibinfo{author}{\bibfnamefont{Y.}~\bibnamefont{Moritomo}}, \bibnamefont{and}
  \bibinfo{author}{\bibfnamefont{Y.}~\bibnamefont{Tokura}},
  \bibinfo{journal}{Phys. Rev. B} \textbf{\bibinfo{volume}{53}},
  \bibinfo{pages}{14285} (\bibinfo{year}{1996}).

\bibitem[{pai()}]{pair}
\emph{\bibinfo{title}{In order to compare the {J}'s in the {FMM} state with
  those in the {CE}-type phase one needs to divide the latter by a factor
  two.}}

\end{thebibliography}
\end{document}